\documentclass[preprint,3p,12pt]{elsarticle}
\usepackage{libertine}
\emergencystretch=\maxdimen
\usepackage{natbib}
\biboptions{sort&compress}
\graphicspath{{Image/}}    
\usepackage[colorlinks]{hyperref}
\AtBeginDocument{%
	\hypersetup{
		plainpages = false, 
		bookmarksopen = true,
		bookmarksnumbered = true,
		breaklinks = true,
		linktocpage,
		colorlinks = true,
		linkcolor = purple,
		urlcolor  = black,
		citecolor = Magenta,
}}

\usepackage{amssymb}
\usepackage{amsmath}
\usepackage{tabularx}
\usepackage{array}
\usepackage{multirow}
\usepackage{booktabs}
\usepackage{adjustbox}
\usepackage{float}
\usepackage{nomencl}
\usepackage[normalem]{ulem}
\usepackage[nodisplayskipstretch]{setspace}
\usepackage{color,soul}
\usepackage[usenames,dvipsnames]{xcolor}
\usepackage{amsfonts,amsmath,amssymb,gensymb}
\usepackage{latexsym,array}
\usepackage{textcomp}
\usepackage{etoolbox}
\makenomenclature
\renewcommand{\nomgroup}[1]{%
  \ifthenelse{\equal{#1}{G}}{\item[\textbf{Greek Symbols}]}{%
  \ifthenelse{\equal{#1}{M}}{\item[\textbf{Abbreviations}]}{}}}
\journal{..}
\usepackage{easyReview} 
\newcommand{\removeEq}[1]{\ifistoreview{\@\expandafter\removeColor{#1}\hspace{-0.6em}} \else {}\fi}
\begin{document}

\begin{frontmatter}

%
\title{{CFD analysis of the influence of solvent viscosity ratio on the creeping flow of viscoelastic fluid over a channel-confined circular cylinder}}
%
%
%
\author[labela]{Pratyush Kumar Mohanty\corref{labelc}}
%
\author[labela]{Akhilesh Kumar Sahu\corref{corr}} 
\emailauthor{sahuak@nitrkl.ac.in}{A.K. Sahu}
\author[labelb]{Ram Prakash Bharti\corref{corr}}
\emailauthor{rpbharti@iitr.ac.in}{R.P. Bharti}
\address[labela]{Computational Flow Modelling (CFM) Research Lab, Department of Chemical Engineering, National Institute of Technology, Rourkela 769008, Odisha, India}
%
\cortext[labelc]{Undergraduate student (Research intern, CFDM Lab, Chemical Engineering, IIT Roorkee; December 2022)}
\address[labelb]{Complex Fluid Dynamics and Microfluidics (CFDM) Lab, Department of Chemical Engineering, Indian Institute of Technology Roorkee, Roorkee 247667, Uttarakhand, India}
\cortext[corr]{Corresponding author}
\begin{abstract}
	\fontsize{11}{21pt}\selectfont
	In this study, the role of solvent viscosity ratio ($\beta$) on the creeping flow characteristics of Oldroyd-B fluid  over a channel-confined circular cylinder has been explored numerically. The hydrodynamic model equations have been solved by RheoTool, an open-source toolbox based on OpenFOAM, employing the finite volume method for extensive ranges of  Deborah number ($De = 0.025-1.5$) and solvent viscosity ratio ($\beta = 0.1-0.9$) for the fixed wall blockage ($B = 0.5$). The present investigation has undergone extensive validation, with available literature under specific limited conditions, before obtaining detailed results for the relevant flow phenomena such as streamline, pressure and stress contour profiles, pressure coefficient ($C_p$), wall shear stress ($\tau_w$), normal stress ($\tau_{xx}$), first normal stress difference ($N_{1}$), and drag coefficient ($C_{\text{D}}$).
	The flow profiles have exhibited a distinctive behavior characterized by a loss of symmetry in the presence of pronounced viscoelastic and polymeric effects.  The results for low $De$ notably align closely with those for Newtonian fluids, and the drag coefficient ($C_D$) remains relatively constant regardless of $\beta$, as the viscoelastic influence is somewhat subdued. As $De$ increases, the influence of viscoelasticity becomes more pronounced, while a decrease in $\beta$ leads to an escalation in polymeric effects; an increase in the $C_D$ value is observed as $\beta$ increases. Within this parameter range, the prevailing force governing the flow is the pressure drag force.  
\end{abstract}
\end{frontmatter}
\section{Introduction}
\noindent
The flow over a cylinder is represented as a conventional bluff body dynamics problem because of its essential nature and vast applicability \citep{streeter1961,Johnson2016}. Despite a simple geometry, this forms the basis for understanding the underlying physics behind flow past complex geometries. It has several industrial uses, such as cooling towers, nuclear reactors, chimney stacks, heat exchangers of the pin and tube types, flow control, and drag reduction techniques. A significant knowledge framework has been established using analytical, experimental, and computational approaches over the years, and numerous studies have explored multiple facets of fluid flow past circular cylinders in static and rotating conditions for both Newtonian and non-Newtonian fluids \citep{GHOSH1994,Zapryanov_1999,zdravkovich1997,zdravkovich2003,Clift_2005,Michaelides_2006,CHHABRA2011aht,Chhabra2023,Loth_2023}.
\\\noindent 
Intricate non-Newtonian characteristics, such as yield stress, shear-rate dependence of viscosity, and viscoelasticity, are commonly recognized in fluids that contain high molecular weight polymers \citep{bird1987,chhabra2011}. Viscoelastic fluids possess the ability to both store and release energy over time, as they exhibit both viscous and elastic behaviour. In industrial operations, the properties of such fluids can substantially affect momentum transfer. Due to the complexity and difficulties in anticipating the viscoelastic fluid behaviour, plenty of viscoelastic fluid phenomena still need to be discovered. 
\\\noindent 
Researchers have employed various experimental and computational methods to comprehend the detailed behavior of viscoelastic fluids and their effect on momentum transfer. Understanding the effects of shear rate, pressure, and temperature on fluid behaviour and devising models that can accurately predict the behaviour of these fluids in industrial contexts are among the areas of emphasis \citep{liu1998,mckinley1993}. Despite the challenges associated with studying viscoelastic fluids, there have been significant advances in the past several years. Some widely used viscoelastic models are Kelvin, Maxwell, FENE-P, Oldroyd-B. Our investigation uses the Oldroyd-B constitutive model, which accurately captures the rheological behavior of the viscoelastic fluids.  The adoption of this model is justified due to its simplicity, based on a sole conformation tensor and two parameters related to relaxation time and polymer concentration \citep{hamid_sasmal_2023}. 
There is a voluminous literature on the fluid flow behaviour of an Oldroyd-B fluid over a circular cylinder at a constant solvent viscosity ratio ($\beta$). However, none of the studies depicts the role of solvent viscosity ratio ($\beta$) on the fluid flow characteristics over a confined cylinder. Therefore, the purpose of this research is to address the existing gap in the literature. Nevertheless, a concise overview of the existing knowledge on Newtonian and non-Newtonian fluids flow over a cylinder is presented to conceptualize the significance of the present study.
\section{Literature Review}
\noindent 
Over the century, the flow over a cylinder has been one of the most classically researched problems to understand the hydrodynamics of bluff bodies, as evidenced by several excellent articles and books \citep{GHOSH1994,Williamson_1996,Zapryanov_1999,zdravkovich1997,zdravkovich2003,Clift_2005,Michaelides_2006,Choi_2008,CHHABRA2011aht,Chhabra2023,Loth_2023} featuring the varieties of the flow characteristics of both Newtonian (and non-Newtonian) fluids around unconfined (and channel-confined) cylinders.  Numerous outstanding studies have demonstrated the influence of wall blockage on the fluid flow characteristics of a cylinder using different means of analysis \citep{Chhabra2023,Loth_2023}. In this section, the literature focusing on Newtonian and non-Newtonian fluid flow around a confined circular cylinder has been reviewed, followed by the detailed literature analysis for Oldroyd-B fluid flow past the cylinder.
\\\noindent
\citet{zdravkovich1997,zdravkovich2003} comprehended the existing knowledge to present the fundamental aspect of the Newtonian flow over a cylinder, including the detailed characterization of flow regimes and highlighted the crucial flow kinematics such as flow transition and wake separation. 
\citet{zovatto2001} conducted in-depth numerical analysis on the impact of wall confinement on two-dimensional steady to vortex shedding flow of Newtonian fluid over a cylinder using the vorticity-stream function formulation and finite element method. 
In a numerical analysis of the steady laminar flow ($Re= 0.1 - 200$) of a Newtonian fluid around a circular cylinder situated within a planar rectangular channel (blockage ratio, $B=H/D = 1.54 - 20$), \citet{chakraborty2004} reported a reduction in drag coefficient $(C_D)$ with increasing $B$ (for fixed $Re$) and increasing $Re$ (for fixed $B$). Further, both separation angle and wake length increased with increasing $Re$ (for fixed $B$). \citet{mettu2006} numerically evaluated the forced convection from an isothermal cylinder confined asymmetrically (gap ratio, $G = 0.125 - 1$; $G=1$ for symmetric) in a planar channel for $Re=10 - 500$, $Pr=0.744$, $B=2.5 - 10$. They observed an increase in critical $Re$ for transiting steady to unsteady flow, an increase in both drag ($C_D$) and Strouhal number ($St$), and a negligible influence on Nusselt number ($Nu$) with decreasing $G$ for all values of $B$. For low $Re$ flow over a confined cylinder, \citet{singha2010} confirmed delayed transition in the vortex shedding with decreasing $B$ due to the strengthening interaction between wake and channel walls and stated that $St$ exhibited independence from $Re$ for low $B$. \citet{sahin2004} performed linear stability analysis to assess the stability of the steady asymmetric solutions for the Newtonian flow ($0< Re <280$) around a channel-confined ($0.1<B<0.9$) cylinder. They observed that the asymmetric flows become unstable as well, transitioning to unsteadiness through a Hopf bifurcation for $B>0.82$. 
Subsequently, \citet{mishra2021} outlined that the presence of large confinement (B = 0.9) stabilizes the 2-D flow ($Re = 4 - 100$) over a cylinder. The confinement delays the start of laminar separation until $Re = 27.8$. Additionally, the steady-state flow was maintained up to $Re = 100$. They also observed that the sudden decrease in surface pressure around the cylinder and the delayed appearance of an abrupt pressure gradient across the cylinder surface are direct outcomes of substantial blockage. 
\\\noindent 
Further, existing literature has significantly explored the flow of power-law fluids around both circular and non-circular cylinders. For instance, the drag and heat transfer characteristics are reported to be complex \citep{bharti2007,bharti2007b} for the steady power-law flow ($1\le Re\le 40$, $0.2\le n\le 1.9$, $1\le Pr\le 100$) across the channel-confined ($1.1\le B\le 4$) cylinder. Similarly, unsteady power-law fluid flow around a confined circular cylinder has been investigated under a broad range of parameters $0.4<n<1.8$, $40<Re<140$,  $B$ = 2, 4, and 6 \citep{rao2011} and for $0.4<n<1.8$, $50<Re<150$,  $B$ = 4 \citep{bijjam2012,bijjam2015}. \citet{vishal2021} determined the critical parameters for shear-thickening ($1<n<1.8$) power-law fluid flow through a channel-confined circular cylinder for two wall blockage ratio values ($B = 2$ and $4)$ using open-source finite volume solver OpenFOAM, and supplemented the results for the critical parameters for unconfined ($B = \infty$) power-law fluid flow \citep{sivakumar2006} obtained using commercial finite volume solver Ansys FLUENT.
\\\noindent
In contrast, the studies related to the Oldroyd-B fluid flowing over a confined cylinder are scant compared to its Newtonian and power-law counterparts.  Most of the investigations have considered a constant solvent viscosity ratio ($\beta = 0.59$, i.e., Boger fluid) for a fixed blockage ratio ($B =2$) and Deborah number ranging from 0 to 2 \citep{caola2001, owens2002, kim2004, hulsen2005, carrozza2019}.
\citet{oliveira1998} developed a new finite volume methodology for the computation of the viscoelastic fluid flow. The proposed approach for acquiring stress values at cell faces was demonstrated effectively, as it successfully eradicates oscillations in the calculated profiles for various quantities even at high $De$. This technique has subsequently been improved \citep{alves2001} for the Oldroyd-B and upper convected Maxwell (UCM) fluids by employing two high-resolution methods, namely, SMART and MINMOD. The numerical accuracy of the improved technique was ascertained in the literature \citep{caola2001,kim2004,owens2002} by comparing the drag and lift coefficient ($C_D$ and $C_L$) values for the benchmark channel confined ($B=0.5$) cylinder flow problem. 
\citet{dou2007} adopted the Oldroyd-B model to compute the confined flow around a circular cylinder. They observed decreasing drag coefficient at low $De$ ($\le 0.6$); however, it increases for higher $De$ as $De$ increases.  Additionally, they demonstrated that the pressure distortion near the cylinder originates from the interplay of normal stress and the influence of streamline curvature, inducing an inflection in the velocity profile. Consequently, induced flow distortion contributes to the overall instability of the flow field. 
\citet{Richter_2010} numerically illustrated the significant influence of dilute polymer additives on the three-dimensional inertial flow of viscoelastic (L2 in the FENE-P model) fluid past a cylinder at $Re=100$ and 300. Recently, \citet{Minaeian_2019,Minaeian_2022} have numerically analyzed viscoelastic effects on the onset of vortex shedding over a channel confined ($B=0.05$) circular cylinder for a high concentration polymer solution governed by the Phan–Thien–Tanner (PTT) model for wide range of elasticity number ($El = 0 - 100$) at $Re=100$ using the rheoFoam - OpenFOAM solver based on the finite volume method. \citet{Hopkins2022} reported novel inertia-less, shear-thinning viscoelastic flow instability for wormlike micellar solution flowing past a confined ($B=0.5$) microcylinder for broad Weissenberg number ($0.5\le Wi\le 900$). \citet{Kumar2022} have numerically explored the hysteresis in pulsatile viscoelastic (FENE-P model) flow instability of confined cylinders using RheoTool integrated with openFOAM and log-confirmation approach for polymeric stress tensor for small Reynolds number ($Re=0.0004-0.004$), constant elasticity number ($El =781.25$), and range of Weissenberg number ($0\le Wi\le 4$).
\\\noindent 
Based on the above discussion of the existing relevant literature, it is notable that limited studies have focused on the viscoelastic (Oldroyd-B) fluid flowing past a confined circular cylinder. To the best of our knowledge, prior studies have not analyzed the impact of solvent viscosity ratio on the flow characteristics of viscoelastic fluid across a channel-confined circular cylinder. Therefore, current work aims to understand the role of the solvent viscosity ratio ($\beta$) and Deborah number ($De$) on the viscoelastic fluid flow characteristics across a confined circular cylinder in the creeping flow regime using finite volume method open-source solved RheoTool integrated with openFOAM and log-confirmation approach for polymeric stress tensor.
%
\section{Problem Formulation}
\begin{figure}[!b]
\centering
\includegraphics[width=\textwidth]{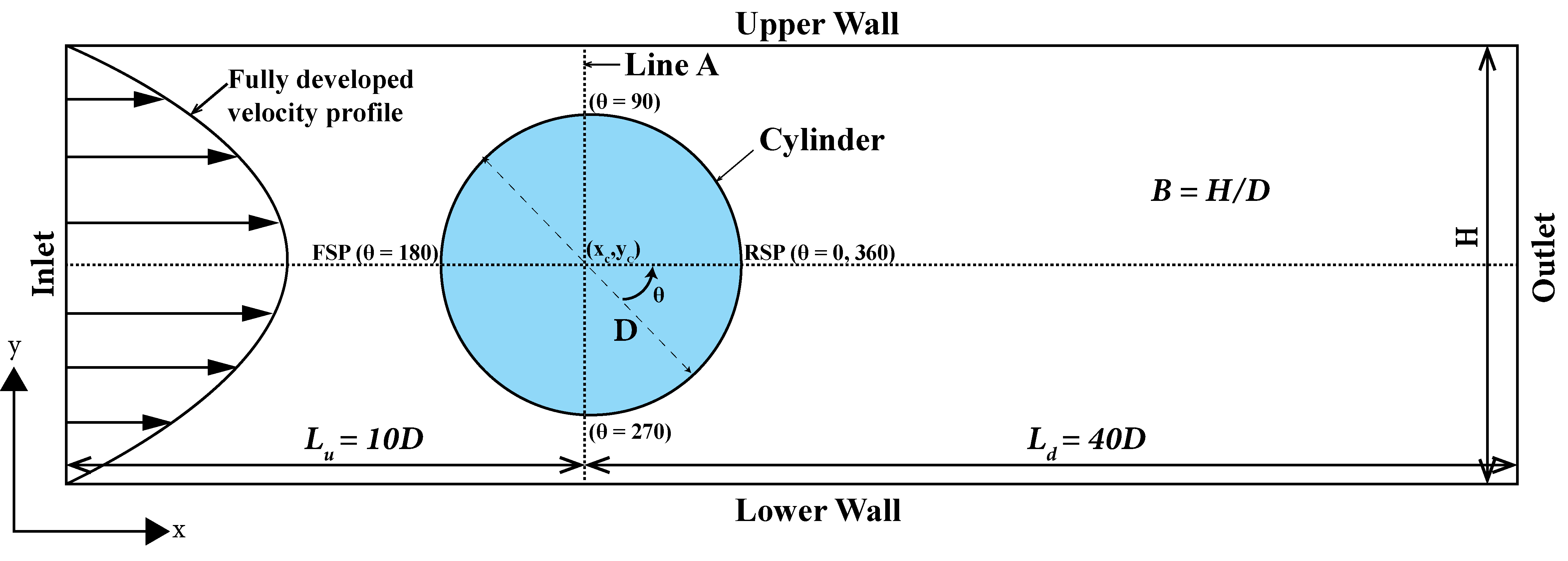}
\caption{Schematic representation of the computational flow arrangement.}
\label{flow_domain}
\end{figure}

%
\noindent
The current study examines the flow properties of a viscoelastic fluid in a two-dimensional (2--D) laminar creeping flow past a circular cylinder (diameter $D$), confined within the two parallel walls spaced apart by a distance $H$, as shown schematically in Fig. \ref{flow_domain}. The inlet flow is considered to have a fully developed velocity profile with a maximum velocity of $U_\text{max}$ and an average velocity of $U_\text{avg}=(2/3)U_\text{max}$. In order to preclude any three-dimensional (3-D) effects, the cylinder is taken to be infinitely long along the z-axis. The wall blockage ratio ($B$) is defined as $B=D/H$. The cylinder is horizontally positioned with center ($x_c$, $y_c$) at ($L_u$, H/2), where $L_u$ is the upstream length measured from the inlet to the center of the cylinder.
The geometrical specifications (length and height) of the rectangular computational flow domain are $L (=L_u+L_d)$ and $H$, where $L_d$ is the downstream length measure from the center of the cylinder to outlet. Further,  $\theta = 0^\circ$ (or $360^\circ$) represents the rear stagnation point (RSP), and $\theta = 180^\circ$ indicates the front stagnation point (FSP) on the surface of the cylinder. 
\\\noindent 
\citet{oldroyd1950} marked the initial attempt to establish constitutive models for viscoelastic fluids while systematically upholding material frame indifference. According to this theory, stress in a continuous media should only be caused by deformations and should not be affected by simple rotation of the material. In this study, the rheological nature of the viscoelastic fluid is defined by the Oldroyd-B model because it adequately approximates a Boger fluid by having a quadratic first normal stress difference, constant shear viscosity, and zero second normal stress difference. 
\\\noindent 
The flow problem is mathematically governed by the continuity and momentum equations in conjunction with the Oldroyd-B viscoelastic model as follows.
\begin{gather}
\frac{\partial {{u_x}}}{\partial x}+\frac{\partial {{u_y}}}{\partial y}=0
\qquad\text{(continuity equation)}
\\
\rho\left[\frac{\partial {u_x}}{\partial t}+\frac{\partial ({u_xu_x})}{\partial x}+\frac{\partial ({u_xu_y})}{\partial y}\right]=-\frac{\partial p}{\partial x}+\left[\frac{\partial {\tau_{xx}}}{\partial x}+\frac{\partial {\tau_{xy}}}{\partial y}\right]
\qquad\text{(x-momentum equation)}
\\
\rho\left[\frac{\partial {u_y}}{\partial t}+\frac{\partial ({u_yu_x})}{\partial x}+\frac{\partial ({u_yu_y})}{\partial y}\right]=-\frac{\partial p}{\partial y}+\left[\frac{\partial {\tau_{yx}}}{\partial x}+\frac{\partial {\tau_{yy}}}{\partial y}\right]
\qquad\text{(y-momentum equation)}
\end{gather} 
In the equations expressed above, ${u_x}$, and ${u_y}$ represents the $x-$ and $y-$ components of velocity vector ($\boldsymbol{U}$), $\rho$ is the density of the fluid, and $p$ denotes pressure. The extra-stress tensor ($\tau$) is a combination of stress due to solvent ($\boldsymbol{\tau}_s$)  and stress due to polymer ($\boldsymbol{\tau}_p$) as follows.              
\begin{equation}
\boldsymbol{\tau}=\boldsymbol{\tau}_s+\boldsymbol{\tau}_p
\end{equation}
The stress due to solvent is expressed as follows:
\begin{equation}
\boldsymbol{\tau}_s=2\eta_s\boldsymbol{D},
\qquad\text{where}\qquad 
\boldsymbol{D} = \frac{1}{2}\left[(\nabla \boldsymbol{U})+(\nabla \boldsymbol{U})^T\right]
\end{equation}
where, $\boldsymbol{D}$ is the deformation rate tensor and $\eta_s$ is the viscosity of the solvent.
\\\noindent
The polymeric stress tensor ($\boldsymbol{\tau}_p$) is computed by the log-conformation method using the non-dimensional configuration tensor ($\boldsymbol{A}$), in accordance with the Oldroyd-B model as follows:
\begin{equation}
\boldsymbol{\tau}_p=\frac{\eta_p}{\lambda}(\boldsymbol{A}-\boldsymbol{I})
\end{equation}
\begin{equation}
\overset{\rm \nabla}{\rm \boldsymbol{A}}=\frac{1}{\lambda}(\boldsymbol{A}-\boldsymbol{I})
\end{equation}
 In the above equations, $\eta_p$ is the polymeric viscosity, $\lambda$ denotes the relaxation time, $\boldsymbol{I}$ represents an identity tensor, and $\overset{\rm \nabla}{\rm \boldsymbol{A}}$ is Oldroyd derivative which is given by
\begin{equation}
\overset{\rm \nabla}{\rm \boldsymbol{A}}=\frac{\partial \boldsymbol{A}}{\partial t}+\boldsymbol{U}\cdot \nabla \boldsymbol{A}-(\nabla \boldsymbol{U})^T\cdot \boldsymbol{A}-\boldsymbol{A}\cdot (\nabla \boldsymbol{U})
\end{equation}
The boundary conditions for the current flow problem are written as follows:
\begin{itemize}
\item\textit{At the inlet boundary:} Fluid enters the computational domain with a fully developed velocity field at the inlet ($0\le y\le H$). In addition, a pressure gradient and the excess stress tensor are ascribed to a value of zero. Mathematically, it is expressed as follows.
\begin{equation}
{u_x}= {U}_\text{avg}\left(1-\left|1-\frac{2y}{H}\right|^2\right),
\qquad 
{u_y}=0, 
\qquad 
\frac{\partial p}{\partial x} = 0
\qquad\text{and}\qquad
\boldsymbol{\tau} = 0
\end{equation}
\item\textit{On the cylinder surface, upper and lower walls:} 
The no-slip velocity condition is used, and the gradient of the pressure is taken to be zero, mathematically expressed as follows. Further, the extra stress tensor is calculated using linear extrapolation.
\begin{equation}
	{u_x}=0,
	\qquad 
	{u_y}=0, 
	\qquad\text{and}\qquad
	\frac{\partial p}{\partial n} = 0
\end{equation}
where, $n$ refers to the direction normal to the boundary.
\item\textit{At the outlet boundary:} The gradient of velocity is assigned a zero value means there is no diffusion flux along the direction normal to the outlet and pressure is ambient.
\begin{equation}
\frac{\partial {u_x}}{\partial x}=0,
\qquad
\frac{\partial {u_y}}{\partial x}=0
\qquad\text{and}\qquad
p = 0
\end{equation}
\end{itemize}
In this work, the dimensionless groups governing the considered flow problem are defined as follows, based on the scaling of the velocity field, pressure field and the stress components with the average velocity (${U}_\text{avg}$), dynamic pressure ($\frac{1}{2}\rho U^2_\text{avg}$) and dynamic stress ($\tau_0= {\eta_0 U_\text{avg}}/{D}$), respectively.
\begin{itemize}
\item Solvent viscosity ratio $(\beta)$ is expressed as follows.
\begin{equation}
	\beta=\frac{\eta_s}{\eta_0}=\frac{\eta_s}{\eta_s+\eta_p}
\end{equation}
where, $\eta_0$ is the total viscosity of the viscoelastic material.
\item Reynolds number $(Re)$ relating inertial to viscous force is given as follows.
\begin{equation}
Re=\frac{\rho {U}_\text{avg} D}{\eta_0}
\end{equation}
\item Deborah number $(De)$ characterizing the fluidity of viscoelastic material relates the relaxation time of the material to the characteristic time scale as follows.
\begin{equation}
De=\frac{\lambda {U}_\text{avg}}{D}
\end{equation}
where, $\lambda$ is the relaxation time; it is the characteristic property of the material \citep{barnes1989}.
\end{itemize}
Furthermore,  the flow characteristics and engineering parameters deduced from the numerically obtained flow fields are defined as follows. The total drag coefficient ($C_{\text{D}}$) over the surface of a cylinder \citep{bharti2007} is evaluated using the following expression.
\begin{equation}
	C_{\text{D}}=\frac{F_{\text{D}}}{\eta_0{U}_\text{avg}}=\frac{1}{\eta_0{U}_\text{avg}}{\underset S \int}(-p\boldsymbol{I}+ \boldsymbol{\tau})\cdot \boldsymbol{\hat{i}}\cdot d\textbf{S}
	\label{eq:cd}
\end{equation}
where, $F_{\text{D}}$ is the total drag force acting in the flow direction over the per unit length of the cylinder, $\textbf{S}$ is the surface area of the cylinder, $\textbf{S}$ is identity matrix and $\boldsymbol{\hat{i}}$ is a unit vector in the flow (i.e., $x$-) direction. The two additive terms in Eq. (\ref{eq:cd}) indicating the contribution of pressure and viscous forces are commonly referred to as the pressure drag coefficient ($C_{\text{DP}}$) and viscous drag coefficient ($C_{\text{DF}}$), respectively.
\\\noindent
The magnitude of the wall shear-stress (WSS) over the surface of the cylinder is obtained using 
\begin{equation}
	\boldsymbol{\tau}_{\text{w}}=\left|\boldsymbol{\hat{n}}\cdot \boldsymbol{\tau} - \boldsymbol{\hat{n}}(\boldsymbol{\hat{n}}\cdot \boldsymbol{\tau}\cdot \boldsymbol{\hat{n}})\right|
\end{equation}
where $\boldsymbol{\hat{n}}$ is the unit vector normal to the surface of cylinder.
\\\noindent
The pressure coefficient ($C_{\text{p}}$) over the surface of the cylinder \citep{bharti2007} is evaluated as.
\begin{equation}
	C_p=\frac{(p - p_\infty)}{\frac{1}{2}\rho {U}^2_\text{avg}}
	\label{eq:cp}
\end{equation}
where, $p_{\infty}$ is the pressure far away from the cylinder under the fully developed condition.
%
\section{Numerical Methodology}
%
\noindent
In this study, an open-source toolbox, rheoTool v6 \citep{pimenta2017,rheoTool}, based on the finite volume method (FVM) CFD open-source code OpenFOAM v7 \citep{openfoam}, is used to solve viscoelastic flow governing equations described in the previous section. The Oldroyd-B model has been used to govern the rheological nature of the viscoelastic fluid material. The high-resolution CUBISTA (Convergent and Universally Bounded Interpolation Scheme for Treatment of Advection) scheme has been used to discretize the advective components of the momentum and constitutive equations because of its superior convergence features \citep{alves2003}. 
The temporal derivative terms are discretized using the linear interpolation technique, and the diffusive terms in the momentum equation have been discretized using the Gauss linear interpolation scheme. An open-source mesh generator, Gmsh v4.11.1 \citep{Gmsh}, has been used to discretize the computational domain and generate a suitable computational grid structure. The mesh is then imported to OpenFOAM. The solution of linearized algebraic equations has been obtained by interfacing the rheoTool with the sparse matrix solvers of PETSc (Portable, Extensible Toolkit for Scientific Computation) (v 3.15) library \citep{balay2021a,balay2021b} that utilizes the direct preconditioner, PCLU, a direct solver for the linear system that employs LU factorization. The stress field is solved using the PBiCG (Preconditioned Bi-conjugate Gradient) solver along with DILU (Diagonal-based Incomplete LU) preconditioner \citep{ajiz1984,lee2003}.
The SIMPLE (Semi-Implicit Method for Pressure Linked Equations) technique has been used to establish the pressure-velocity coupling and the numerical solution is stabilized using the log-conformation tensor method \citep{pimenta2017,fattal2004}. The semi-coupled solver, wherein pressure and velocity are coupled, and stress is segregated, is used to obtain the fully converged solution (with a relative tolerance of  $10^{-10}$) for the pressure, velocity, and stress fields.
\begin{figure}[!b]
	\centering
	\includegraphics[width=\textwidth]{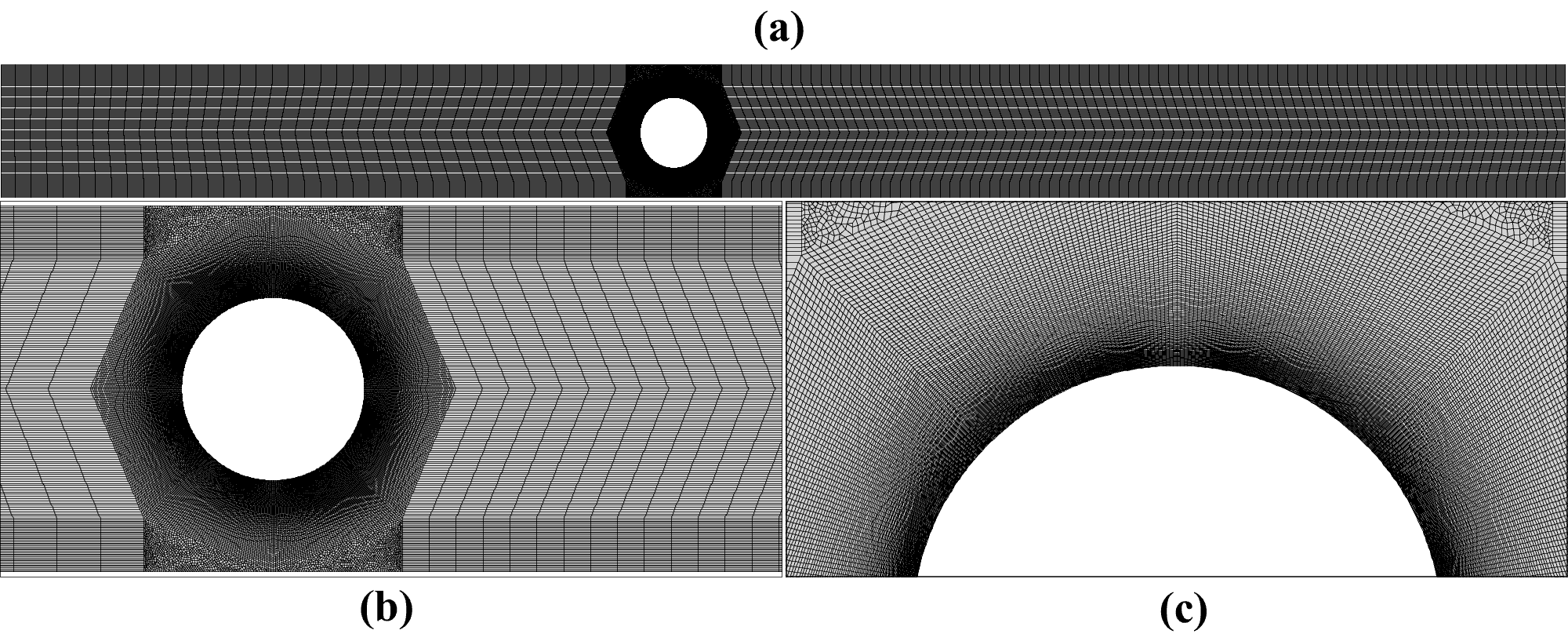}
	\caption{Schematics of mesh (a) in the whole computational domain (b) around the cylinder (c) zoomed view near the cylinder.}
	\label{mesh}
\end{figure}
\\\noindent
The appropriate selection of computational and numerical parameters (such as domain dimensions, grid resolution, and time-step size) is significant to the precision of the obtained results. It is, therefore, important to suitably select these parameters by trade-off the accuracy of results and the least amount of computing time.
\begin{table}[!b]
  \caption{Grid independence test for extreme values of Deborah number ($De=0.025$ and 1.5) and solvent viscosity ratio ($\beta=0.1$ and 0.9).}
  \label{tab:grid_drag}
  \centering
  \begin{adjustbox}{width=\textwidth}
  \begin{tabular}{@{}cccccccc@{}}
    \toprule
     &  & &  & \multicolumn{4}{c}{{Drag Coefficient ($C_{\text{D}}$)}} \\
     \cmidrule(lr){5-8}
    {Grid} & {$N_{\text{c}}$} & {$N_{\text{n}}$} & {$N_{\text{e}}$} & \multicolumn{1}{c}{{$De=0.025$, $\beta=0.1$}} & \multicolumn{1}{c}{{$De=0.025$, $\beta=0.9$}} & \multicolumn{1}{c}{{$De=1.5$, $\beta=0.1$}} & \multicolumn{1}{c}{{$De=1.5$, $\beta=0.9$}} \\
    \midrule
    G1 & 240 & 129882 & 195493 & 131.0147 & 132.3322 & 115.5089 & 140.4210 \\
    G2 & 480 & 158202 & 237973 & 131.1744 & 132.6261 & 118.5783 & 143.2001 \\
    G3 & 720 & 186522 & 280453 & 131.2015 & 132.6828 & 118.3904 & 143.8222 \\
    G4 & 960 & 214842 & 322933 & 131.2128 & 132.7055 & 118.6864 & 143.9110 \\
    \bottomrule
  \end{tabular}
  \end{adjustbox}
\end{table}
%
As outlined in the preceding section, the computational domain is characterized by upstream length, $L_u$ and downstream length, $L_d$. Based on the systematic domain independence study and our previous experience \citep{bharti2007}, the optimal values are selected as $L_u= 10D$ and $L_u=30D$ for the fixed blockage ratio ($B=0.5$) to alleviate the end effects considering the imposed fully developed inlet and outlet con
\\\noindent
After selecting the appropriate computational domain ($L_u=10D$, $L_d=30D$, $B=0.5$), an extensive grid independence study is performed to ensure the independence of numerical results on the grid structure resulting from the spatial discretization of the computational domain.   
Four distinct non-uniform unstructured grids G1 to G4 (shown schematically in Figure \ref{mesh})  generated using open-source mesh generator Gmsh v4.11.1 \citep{Gmsh} are used to perform the grid independence study. Table \ref{tab:grid_drag} displays the mesh characteristics ($N_{\text{c}}$ is the number of nodes on the surface of the cylinder, $N_{\text{n}}$ is total number of nodes in the computational domain, and $N_{\text{e}}$ is total number of quadrilateral mesh elements in the computational domain) and the influence of mesh on the drag coefficient ($C_{\text{D}}$) for extreme values of Deborah number ($De=0.025$ and 1.5) and solvent viscosity ratio ($\beta=0.1$ and 0.9).
An analysis of Table \ref{tab:grid_drag} suggests that the $C_{\text{D}}$ values show insignificant change ($< 1\%$) with refinement in the grid from G2 to G3 and G4. Therefore, considering the significant enhancement in the computation efforts with mesh refinement, grid G2 is selected to obtain the new results presented in this work. Furthermore, while the considered problem is time-independent, the time-step of $\Delta t=0.001$s is  selected for the present study, as the openFOAM utilizes false-transient approach to obtain the numerical solution. 
\section{Results and discussion} 
\noindent 
The current investigation has explored the influences of the solvent viscosity ratio ($0.1 \leq\beta\leq0.9$) and Deborah number ($0.025\leq De \leq1.5$) on the momentum transfer characteristics of a channel-confined ($B=0.5$) cylinder submerged in low Reynolds number ($Re=0.01$) creeping flow of viscoelastic fluid rheologically governed by the Oldroyd-B model.  The novelty and significance of the considered parameters for investigation are as follows. For larger $\beta$, the fluid typically shows Newtonian behavior as very little to negligible polymeric components in the polymeric solution. As the value of $\beta$ decreases, it tends to become more polymeric. For example, dilute polymer solutions in water have low $\beta$ (= 0.1 - 0.3), concentrated polymer solutions or suspensions have moderate $\beta$ (= 0.4 - 0.6), whereas hydrogels or suspensions with a high solid particle content have high $\beta$ (= 0.7 - 0.9). Further, a broad range of Deborah numbers ($De$) covers the variety of viscoelastic fluids, from almost pure viscous fluids ($De = 0.025$) to highly viscoelastic fluids ($De = 1.5$). Such an investigation is essential for gaining insights into viscoelastic fluid flow in various scientific and engineering applications \citep{Zhou2020,Alves2021,Kumar2023}. 
In this section, the modeling approach has first been validated to ensure the accuracy and reliability of the new results. Subsequently, the detailed results have been shown to elaborate the influence of the flow governing parameters ($\beta$, $De$) on the flow characteristics (such as streamline, pressure, and the normal stress contour profiles, pressure coefficient, normal and wall shear stress over the cylinder, and velocity profiles, drag coefficient and its components). 
\subsection{Validation}
\noindent
In order to ensure the reliability and accuracy of the new result presented in this work, the numerical modeling approach has been validated by comparing the present results for the drag coefficient ($C_\text{D}$) with previously reported data  \citep{dou1999, fan1999, liu1998, sun1999, alves2001} in Table \ref{tab:drag_comparison}. 
\begin{table}[!b]
  \centering
  \caption{Comparison of the present $C_\text{D}$ values with literature for creeping ($Re\rightarrow 0$) flow.}
  \label{tab:drag_comparison}
  \adjustbox{max width=\textwidth}{%
  \begin{tabular}{@{}ccccccc@{}}
    \toprule
     $De$ & Present study &\citet{dou1999} &\citet{fan1999} &\citet{liu1998} & \citet{sun1999}& \citet{alves2001}\\
    \midrule
    0.025 & 132.0944 &131.5020 & -- & -- & --& --\\
    0.050 & 130.6353 &131.0823 &--  &-- &-- & --\\
    0.100 & 126.9604 &129.7226& 130.3597 &--  & 130.3283 & 130.3069 \\
    0.300 & 126.9605 &123.5148 & 123.1893 & -- & 123.2597 & 123.1504 \\
    0.500 & 118.6448 &120.5768 & 118.8301 & 119.5024 & 119.1053 & 118.8137  \\
    0.700 & 122.8196 &121.1297 & 117.3196 & 118.5398 & 117.8361 & 117.3435  \\
    0.900 & 129.4620 &124.4799 & 117.7996 &--  & 118.5033&117.9064 \\
    1.100 & 137.9596 &133.1570&-- &-- &120.3845 &-- \\
    1.300 & 147.5921 &136.9420 &-- &-- &123.0674 &-- \\
    1.500 & 157.8138 &147.1723 &-- &-- &126.3221 &--\\
    \bottomrule
  \end{tabular}
  }
\end{table}
It is evident from the Table \ref{tab:drag_comparison} that the present results are in good agreement ($<1\%$) with the literature results, particularly within the low Deborah number range ($De < 1.2$).  Further, no oscillations are observed even at high Deborah numbers ($De = 1.5$) in this study. This observation of the current analysis shows strong agreement with the literature \citep{dou1999} data, which reported no oscillations even up to $De = 1.8$.  
However, {the present results do not align well with other literature values, as those studies exhibit oscillations at high Deborah numbers}{other studies \citep{fan1999, liu1998, sun1999, alves2001} exhibit oscillations at high $De$, possibly} due to inadequate mesh refinement, due to which present and literature results are not as closely aligned as with \citep{dou1999}. Keeping note that the considered studies \citep{dou1999, fan1999, liu1998, sun1999, alves2001} have used different modeling and numerical approaches to obtain the numerical solution, the present results are considered to display excellent ($\pm 1\%$) accuracy and reliable for design and engineering of relevant applications. This comparative analysis offers further confidence in the accuracy and reliability of the current modeling approach.
\subsection{Streamline patterns}
\begin{figure}[!b]
\centering
\includegraphics[width=0.85\textwidth]{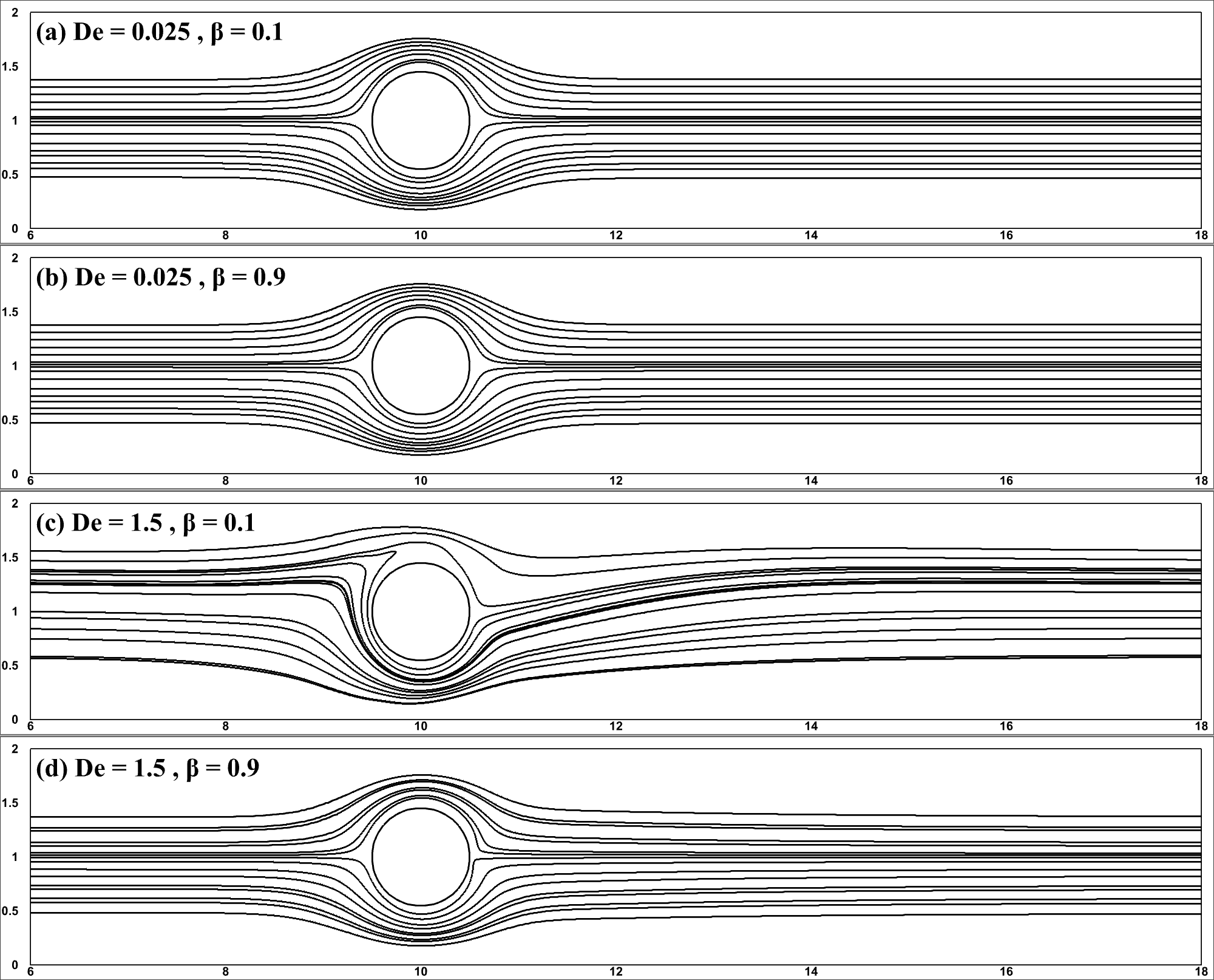}
\caption{Streamline patterns around the cylinder at (a) $De=0.025$, $\beta=0.1$, (b) $De=0.025$, $\beta=0.9$, (c) $De=1.5$, $\beta=0.1$, (d) $De=1.5$, $\beta=0.9$.} 
\label{streamlines}
\end{figure}

\noindent 
In this section, the detailed flow characteristics are presented and analyzed using the streamline ($\psi$) patterns (Fig. \ref{streamlines}) for the extreme values of dimensionless parameters ($\beta$, $De$). Uniformly distributed ($\delta \psi= 0.2$) contour lines ranging from $\psi_{\text{min}}=0$ to $\psi_{\text{max}}=3$ are drawn in Fig. \ref{streamlines}.
As expected, no flow separation behind the cylinder is depicted in Fig. \ref{streamlines} over the range of conditions. 
Viscous force is the main factor causing the flow to be creeping. The viscous force makes the fluid adhere to the surface of the cylinder, thereby preventing any separation or detachment, and flow remains attached.   
At a low value of Deborah numbers ($De=0.025$), the effect of $\beta$ on the streamline patterns is not significant, as observed in Fig. \ref{streamlines}(a, b). 
This phenomenon can be ascribed to the fact that when the Deborah number is low, the viscoelastic effect is diminished due to the shorter relaxation times, resulting in decreased polymeric effects. As a result, the solvent viscosity ratio ($\beta$) has minimal impact on the streamlines for low $De$. However, the streamline profiles at the higher Deborah number ($De = 1.5$) are seen to be asymmetric for lower $\beta$ ($\le 0.2$) and symmetric for $\beta$ ($> 0.2$) about the horizontal centerline ($x, y_c$), as depicted in Fig. \ref{streamlines}(c, d). At high Deborah numbers ($De$), the effects of elastic forces within the fluid become more pronounced. As the ratio of solvent viscosity ($\beta$) decreases, the flow behaviour is increasingly influenced by the elastic properties of the fluid, and the polymeric effects increases. This increased elastic effect due to major polymeric contribution has the potential to cause additional deformation and elongation of the fluid constituents or elements, thereby leading to evident adaptations in the streamline.  
Having observed the complex dependence of the streamline patterns on the Deborah number ($De$) and the solvent viscosity ratio ($\beta$), the subsequent section explores the pressure and stress profiles to gain further insights into the flow.
%
\subsection{Pressure and stress patterns}
\begin{figure}[!b]
	\centering
	\includegraphics[width=0.85\textwidth]{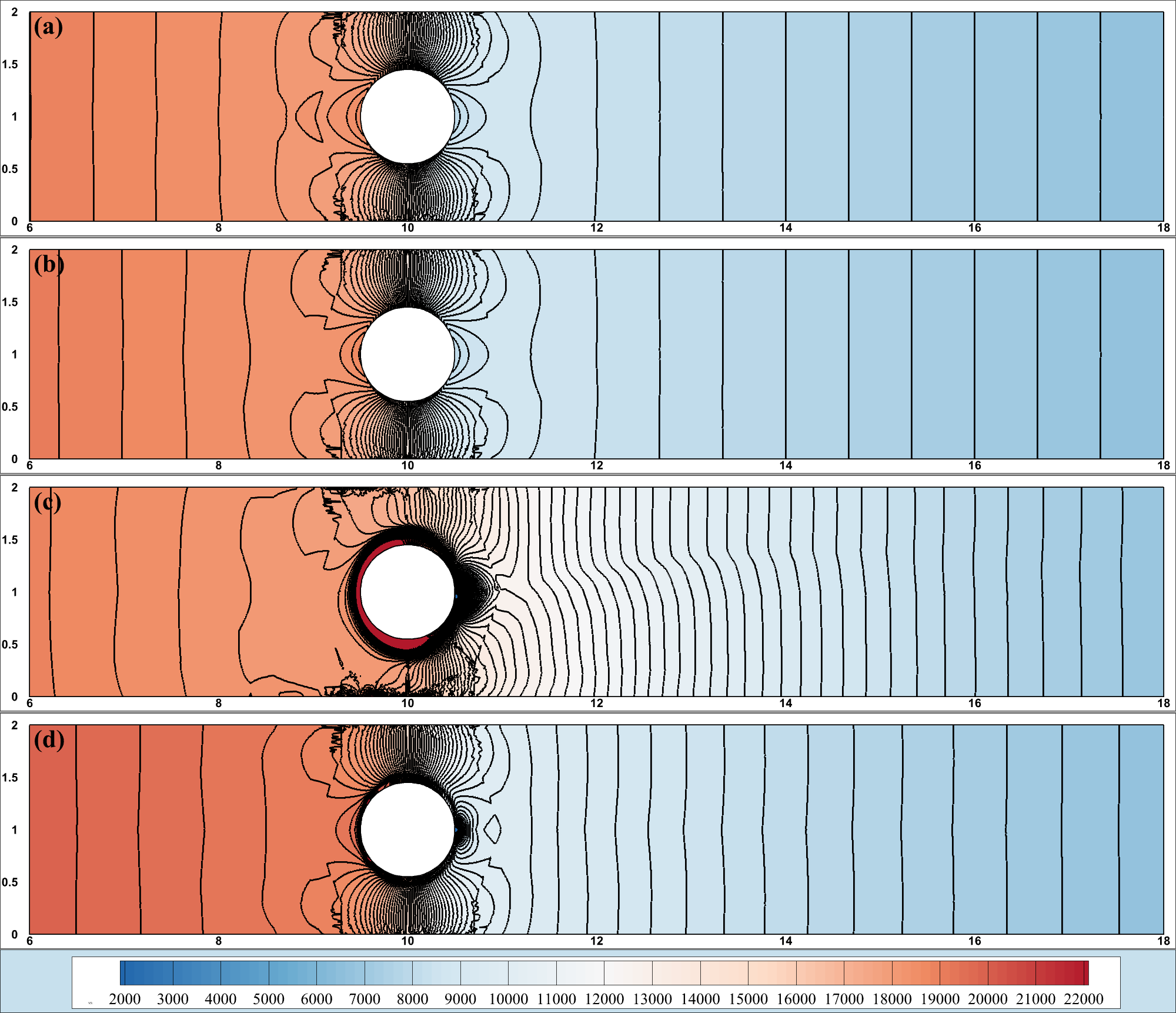}
	\caption{Pressure ($p$) contours around the cylinder for (a) $De=0.025$, $\beta=0.1$ (b) $De=0.025$, $\beta=0.9$ (c) $De=1.5$, $\beta=0.1$ (d) $De=1.5$, $\beta=0.9$}
	\label{pressure_contour}
\end{figure}

\noindent 
Fig. \ref{pressure_contour} illustrates the scaled pressure ($p$) contours around the cylinder for the extreme values of  the Deborah number ($De$) and the solvent viscosity ratio ($\beta$). 
Uniformly distributed ($\delta p= 200$) contour lines ranging from $p_{\text{min}}=2000$ to $p_{\text{max}}=22000$ are drawn in Fig. \ref{pressure_contour}. The variation in dynamic pressure surrounding the cylinder is observed to increase by $30.43 ~\%$  with increasing Deborah number $(De)$ from 0.025 to 1.5. At the low value of $De$, the pressure contours are seen to be symmetric around the cylinder in Fig. \ref{pressure_contour}(a, b). 
\begin{figure}[!b]
	\centering
	\includegraphics[width=0.85\textwidth]{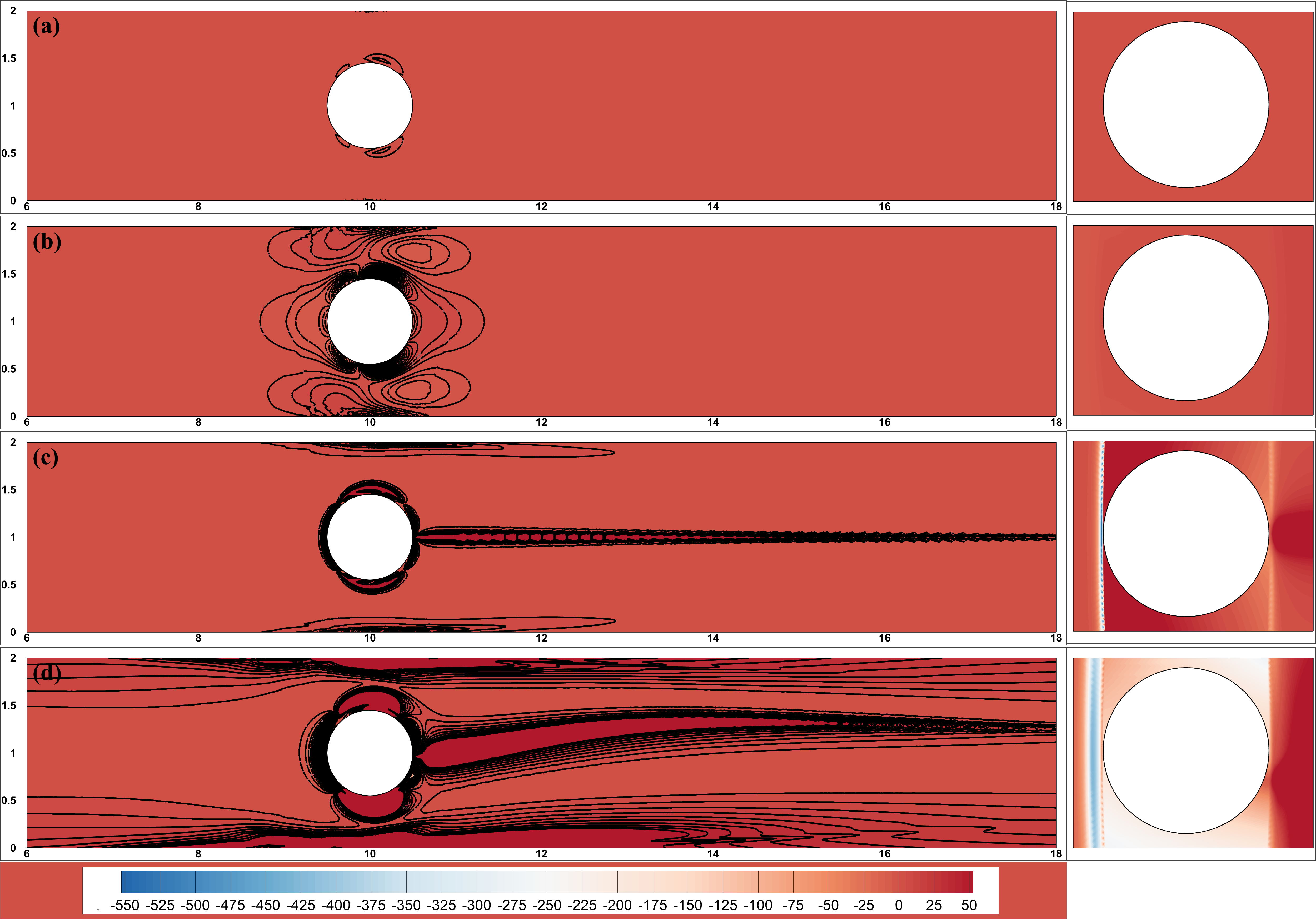}
	\caption{First normal stress difference ($N_{1}=\tau_{xx} - \tau_{yy}$) contours around the cylinder for (a) $De=0.025$, $\beta=0.9$ (b) $De=0.025$, $\beta=0.1$ (c) $De=1.5$, $\beta=0.9$ (d) $De=1.5$, $\beta=0.1$}
	\label{stress_contour}
\end{figure}
There is, however, a pronounced pressure gradient in the vicinity of the cylinder at high $De$, as seen in Figs. \ref{pressure_contour}(c, d). The presence of a thin stress boundary layer on the surface of the cylinder and elongation in the region of stagnation points causes a significant pressure difference around it, as reported in the literature \citep{dou2007}.  The fluid behaviour closely approaches that of a Newtonian fluid at lower values of $De$, and thus the pressure contours show similarity with those seen in Newtonian fluid flow. 
At low Deborah numbers, the pressure exhibits nearly symmetrical contours. However, as the Deborah number increases, the pressure contours undergo distortion in the vicinity of the cylinder. At high $De$ (Fig. \ref{pressure_contour}(c, d)), a protrusion arises at the rear side of the cylinder. As $\beta$ decreases from 0.9 to 0.1, a protrusion that was initially flatter becomes more pointed and enlarges. 
\\\noindent
Subsequently, Fig. \ref{stress_contour} illustrates the scaled first normal stress difference ($N_{1}=\tau_{xx} - \tau_{yy}$) contours around the cylinder for the extreme values of the Deborah number ($De$) and the solvent viscosity ratio ($\beta$). Uniformly distributed ($\delta N_{1}= 5$) contour lines ranging from $N_{1,\text{min}}=-550$ to $N_{1,\text{max}}=50$ are drawn in Fig \ref{stress_contour}.   
The profiles display a complex dependence of the stress on the dimensional parameters ($De$, $\beta$). For instance, in Fig. \ref{stress_contour}(a), where $De$ is low, and $\beta$ is high, there is no induced polymeric effect, and the flow is very similar to that of Newtonian fluid. The stress ($N_1$) contours are only concentrated at the surface of the cylinder. 
\begin{figure}[!b]
	\centering
	\includegraphics[width=\textwidth]{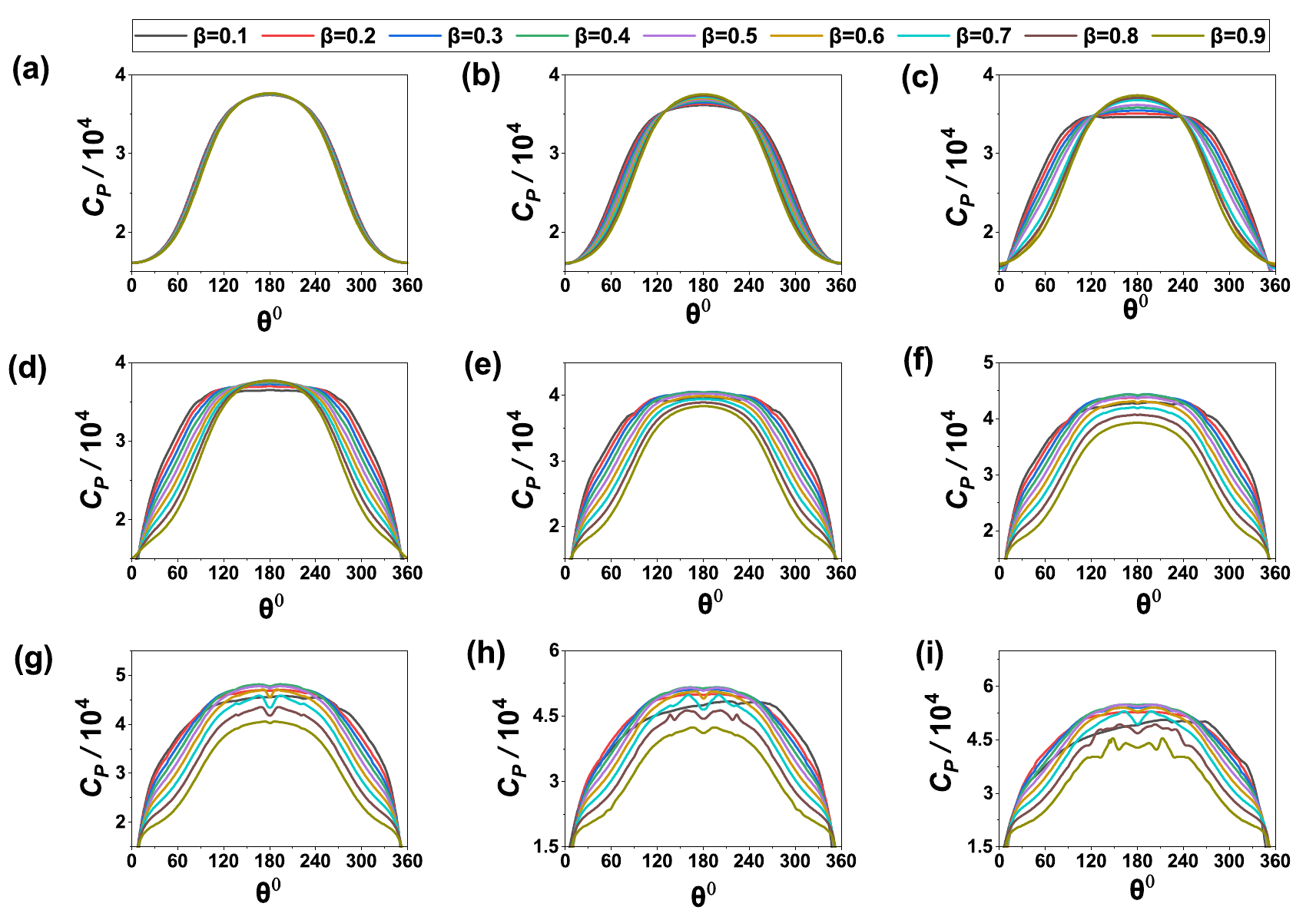}
	\caption{Variation of $C_p$ over the surface ($0^\circ\le \theta\le 360^\circ$) of the cylinder for $0.1\le \beta\le 0.9$ at (a) $De=0.025$ (b) $De=0.1$ (c) $De=0.3$ (d) $De=0.5$ (e) $De=0.7$ (f) $De=0.9$ (g) $De=1.1$ (h) $De=1.3$ (i) $De=1.5$. } 
	\label{Cptheta}
\end{figure}
With increasing $\beta$, at fixed $De$, the polymeric effect tends to induce, and the effects get carried away from the surface of the cylinder to the surrounding regions, as seen in Fig. \ref{stress_contour}(b). Further, with an increase in $De$, refer Fig. \ref{stress_contour}(c,d), the stress wake enlarges and spans almost the whole length of the channel, and a thin stress boundary layer forms on the surface of a cylinder. These observations agree with the results reported \citep{dou2007}. 
%
\subsection{Coefficient of pressure}
%
\noindent 
Fig. \ref{Cptheta} depicts the pressure coefficient ($C_p$, defined in Eq. \ref{eq:cp}) profiles on the surface  ($0^\circ\le \theta\le 360^\circ$) of the cylinder for the considered ranges of $\beta$ and $De$.  Here, $\theta = 180^\circ$ indicates the front stagnation point (FSP), and $\theta = 0^\circ$ (or $360^\circ$) represents the rear stagnation point (RSP) on the surface of the cylinder. The $C_p$ profiles are qualitatively consistent, at the lower values of $De$ and $\beta$, with the literature \citep{Bharti2006,bharti2007} on Newtonian and non-Newtonian power-law fluid flowing across a cylinder. Broadly, $C_p$ profiles are symmetric in the upper and lower half of the cylinder, as expected, due to the creeping nature of the flow. 
For the fixed values of $\beta$ and $De$, $C_p$ increases from the minimum value at RSP ($\theta = 0^\circ$) until it reaches its peak at the FSP ($\theta = 180^\circ$) of the cylinder, and. after that, it starts decreasing until RSP ($\theta = 360^\circ$). 
\begin{figure}[!b]
\centering
\includegraphics[width=\textwidth]{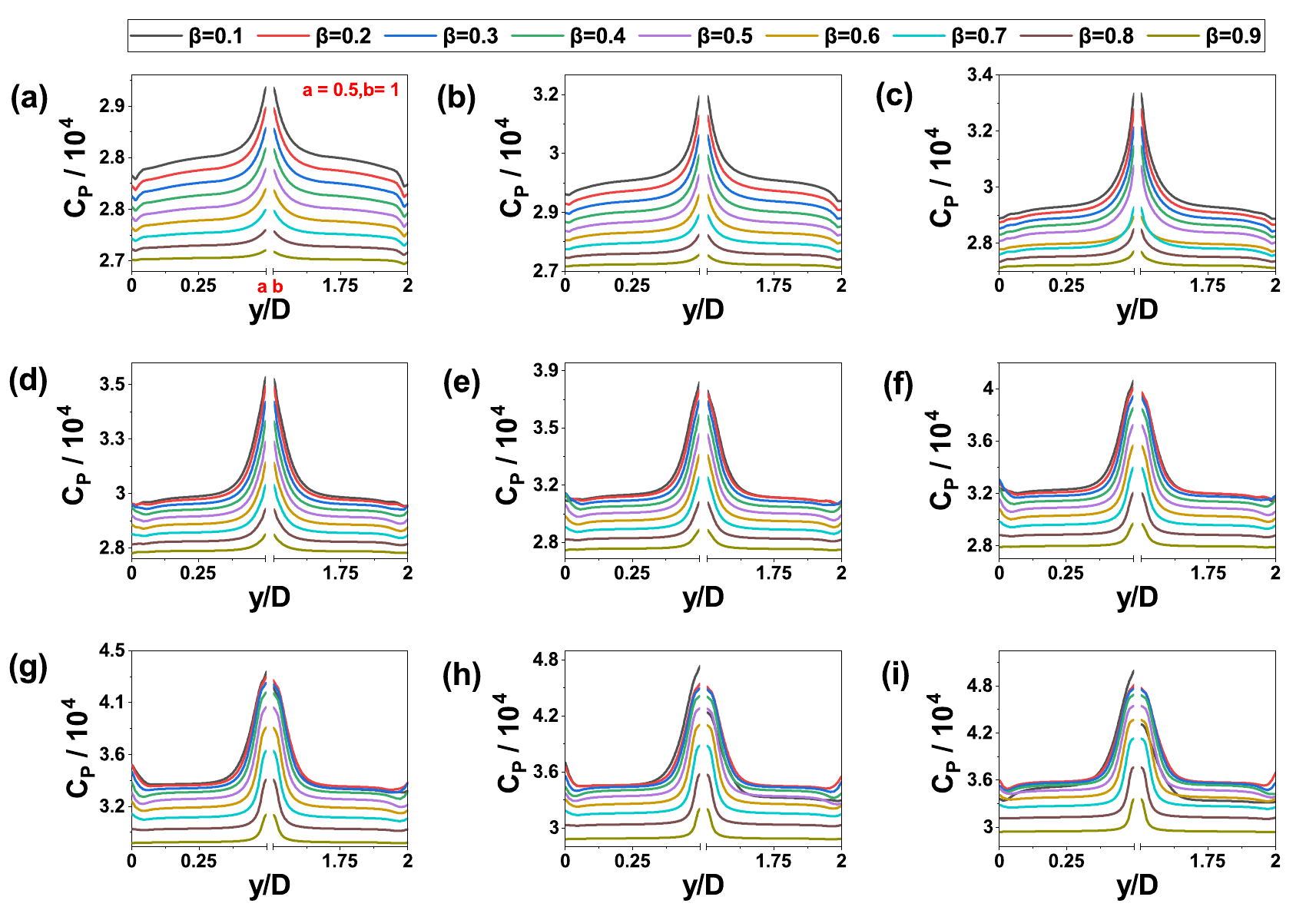}
\caption{Variation of  $C_p$ along line A ($x_c, y$), (refer Fig. \ref{flow_domain}) for different $\beta$ at (a) $De=0.025$ (b) $De=0.1$ (c) $De=0.3$ (d) $De=0.5$ (e) $De=0.7$ (f) $De=0.9$ (g) $De=1.1$ (h) $De=1.3$ (i) $De=1.5$.} 
\label{Cpline}
\end{figure}
For lower values of $De$ and $\beta$, $C_p$ decreases in the frontal side of the cylinder, and subsequently, the curve flattens in the rear side of the cylinder as the polymeric effects tend to dominate, and the boundary layer thickness may increase.  A thicker boundary layer leads to a slower velocity near the surface of the cylinder. This deceleration of the flow near the frontal side of the cylinder decreases in that region. Due to the above-specified reason, crossing over in $C_p$ profiles is seen in Fig. \ref{Cptheta}(a-d). However, at higher values of $De$, the viscoelastic effects tend to dominate, which affects the symmetry of the $C_p$ profiles, especially for the higher values of $\beta$ due to the additional polymeric effects. Furthermore, it is clearly illustrated in Fig. \ref{Cpline} that the influence of $\beta$ is comparatively less pronounced at lower $De$ but has significant effect at higher $De$.
\\\noindent 
Subsequently, Fig. \ref{Cpline} illustrates $C_p$ profile over a vertical line A ($x_c, y$) that passes through the center of the cylinder (refer Fig. \ref{flow_domain}). The discontinuity shown on the $x$-axis of Fig. \ref{Cpline} denotes the region ($0.5\le y\le 1.5$) overlapped with the cylinder. The profiles clearly show that the value of $C_p$ increases from the channel wall ($y=0$ or $H$) up to the surface ($y=0.5$ and 1.5) of the cylinder.  
Furthermore, $C_p$ values decrease with increasing $\beta$ for a given Deborah number ($De$). Interestingly, $C_p$ assumes greater values for lower $\beta$ at a lower Deborah number ($De\le 0.7$). 
%
\subsection{Wall shear stress (WSS)}
%
\begin{figure}[!b]
\centering
\includegraphics[width=\textwidth]{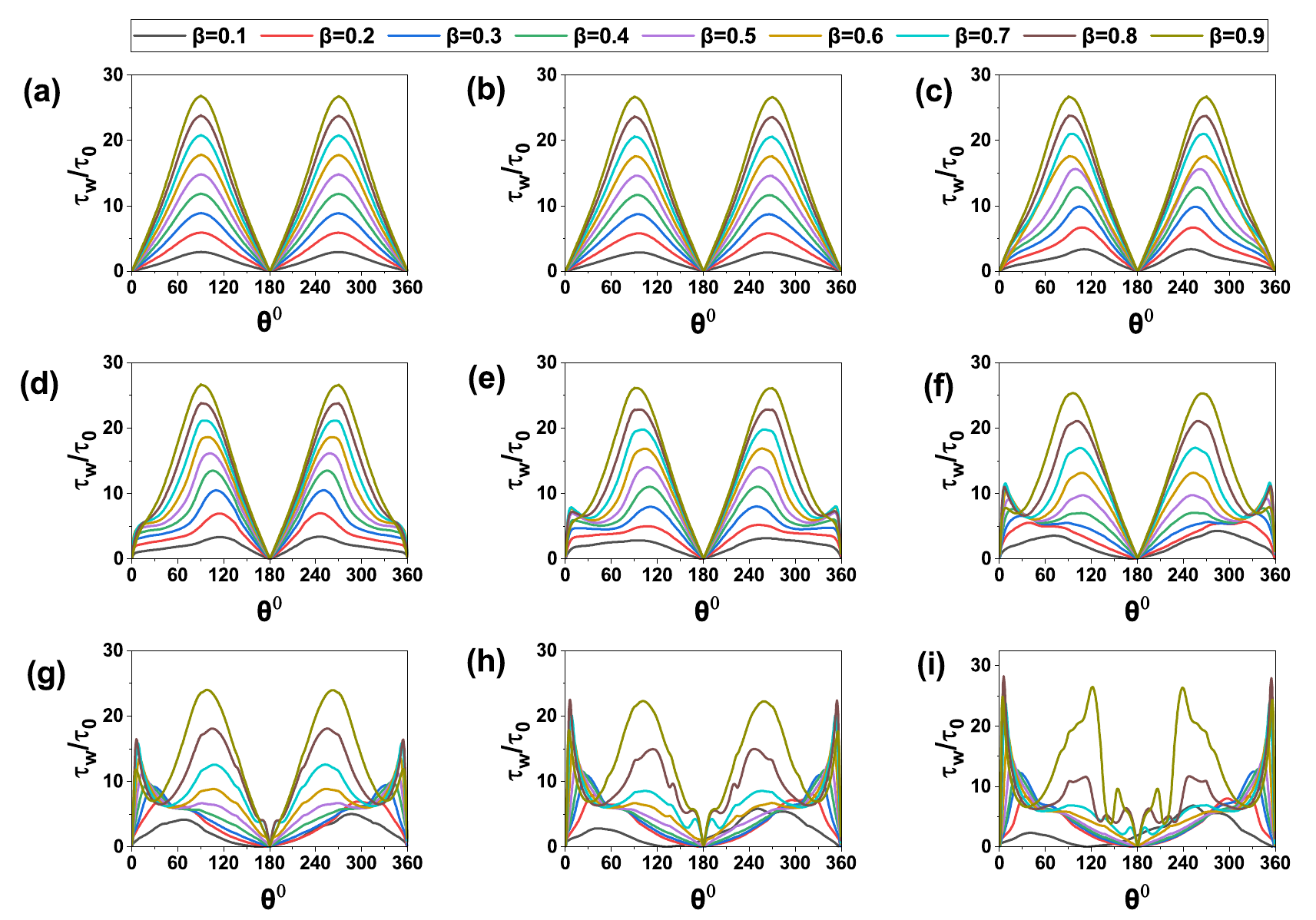}
\caption{Variation of WSS ($|\tau_w/\tau_0|$) over the surface of the cylinder for different $\beta$ at (a) $De=0.025$ (b) $De=0.1$ (c) $De=0.3$ (d) $De=0.5$ (e) $De=0.7$ (f) $De=0.9$ (g) $De=1.1$ (h) $De=1.3$ (i) $De=1.5$.} 
\label{Wsstheta}
\end{figure}
\noindent 
Fig. \ref{Wsstheta} displays the variation of magnitude of the wall shear stress (WSS, $|\tau_w/\tau_0|$) over the surface ($0^\circ\le \theta\le 360^\circ$) of the cylinder over the ranges of considered conditions ($\beta$, $De$).  At both RSP ($\theta = 0^\circ$) and FSP  ($\theta = 0^\circ$ and $360^\circ$), the magnitude of WSS becomes zero (i.e., $|\tau_w/\tau_0|=0$)  due to velocity gradient being zero as the kinetic energy completely converts to the potential energy without any loss of energy at the stagnation point. In general, WSS displays a symmetric nature in the upper ($0^\circ\le \theta\le 180^\circ$) and lower ($180^\circ\le \theta\le 360^\circ$) half of the cylinder. At low $De\le 0.1$, in the upper half ($0^\circ\le \theta\le 180^\circ$), WSS increases upward from RSP to top ($\theta = 90^\circ$), subsequently decreasing up to the FSP; the vice versa is seen in the lower half ($360^\circ\ge \theta\ge 180^\circ$) of the cylinder, as seen in Fig. \ref{Wsstheta}(a, b). WSS has shown the highest and equal magnitude at the top ($\theta = 90^\circ$) and bottom ($\theta = 270^\circ$) of the cylinder, as the velocity gradient gets maximized at these locations due to their close proximity to the channel wall. With increasing $De$, while the symmetry in the profiles remains, the peak of the curves shifts towards FSP, Fig. \ref{Wsstheta}(c, d), and additional minor peaks appear near RSP with decreasing $\beta$, Fig. \ref{Wsstheta}(e, g). The high polymeric effect (i.e., lower $\beta$) attributes in the flattening of WSS profiles for smaller $\beta$ values at high ($De$). There is a sudden spike in the value of WSS at the frontal side of the cylinder at higher $De$ and lower $\beta$, as indicated in Fig. \ref{Wsstheta}(e - i). At higher $De$ and lower values of $\beta$, Fig. \ref{Wsstheta}(g - i), the profiles are no more symmetrical, and the value of WSS is higher on the bottom than at the top of the cylinder.  
One of the potential causes of the phenomena mentioned above could be the start of the wake development at the top side of the cylinder, which may have decreased the WSS there.
\begin{figure}[!b]
\centering
\includegraphics[width=\textwidth]{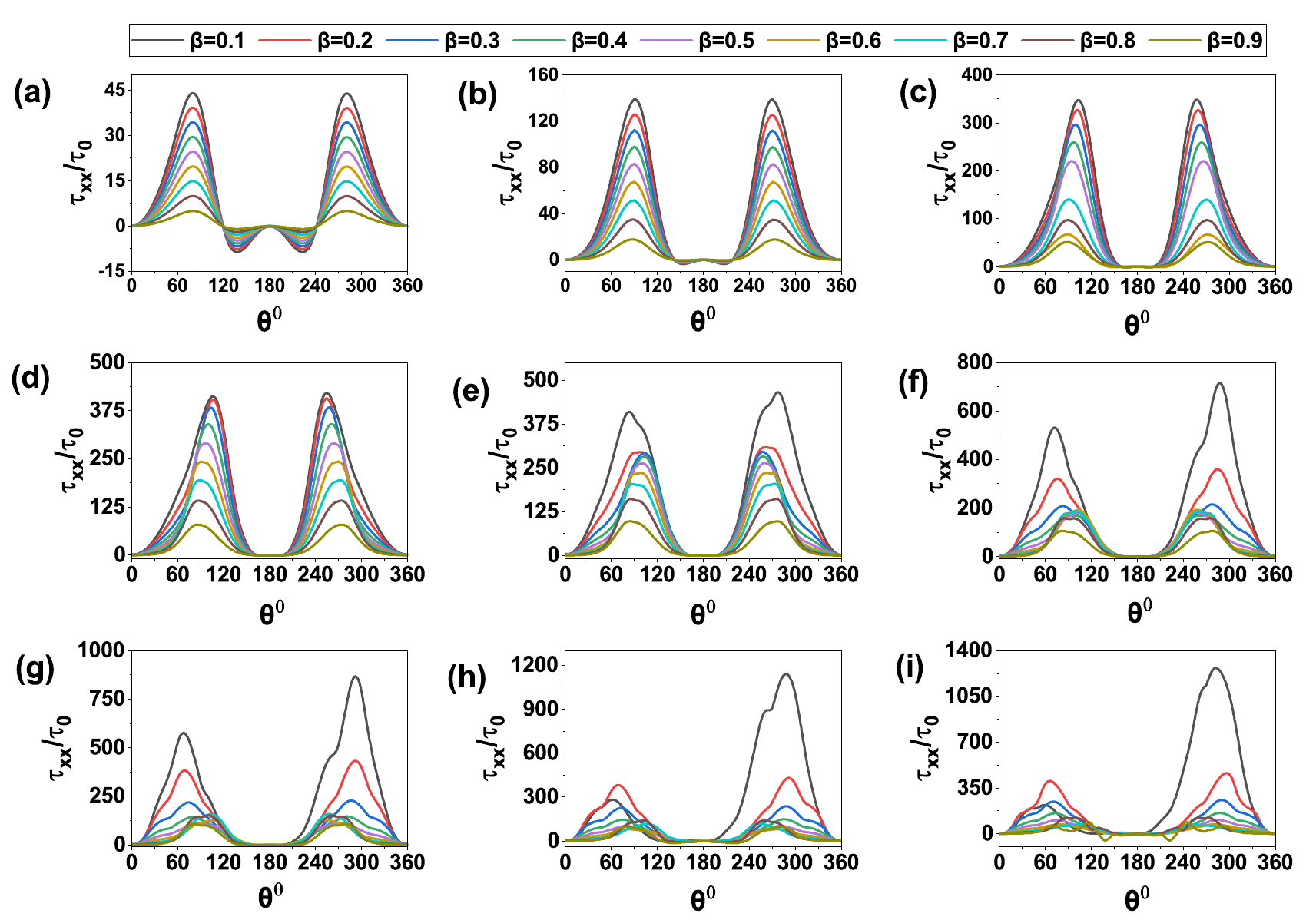}
\caption{Variation of normal component of stress ($\tau_{xx}/\tau_0$) over the surface ($0^\circ\le \theta\le 360^\circ$) of the cylinder for different $\beta$ at (a) $De=0.025$ (b) $De=0.1$ (c) $De=0.3$ (d) $De=0.5$ (e) $De=0.7$ (f) $De=0.9$ (g) $De=1.1$ (h) $De=1.3$ (i) $De=1.5$.} 
\label{tauxxtheta}
\end{figure}
\subsection{Normal Stress}
%
\noindent
Fig. \ref{tauxxtheta} shows the variation of the normal component of stress ($\tau_{xx}/\tau_0$) over the surface ($0^\circ\le \theta\le 360^\circ$) the cylinder. The normal stress has shown significant dependence on the governing parameters ($De$, $\beta$), as compared with  WSS (Fig. \ref{Wsstheta}). The profiles have shown two peaks, each at the top and bottom sides of the cylinder, due to the thinning of the boundary layer. The maximum value of $\tau_{xx}$ increases with $De$ as the viscoelastic effect tends to dominate, and there is an increase in the relaxation time. At a particular $De$, $\tau_{xx}$ increases with a decrease in $\beta$ as the polymeric effects dominate; it increases the induced stress. As expected, $\tau_{xx}$ is zero at FSP and RSP. It can be seen in Fig. \ref{tauxxtheta}(a), $\tau_{xx}$ value deviates from zero and becomes negative on the upper and lower sides of the front portion of the cylinder at lower $De$. However, at higher $De$, $\tau_{xx}$ values are higher on the lower side of the cylinder than the upper side, especially for lower $\beta$.
\begin{figure}[!htb]
\centering
\includegraphics[width=0.75\textwidth]{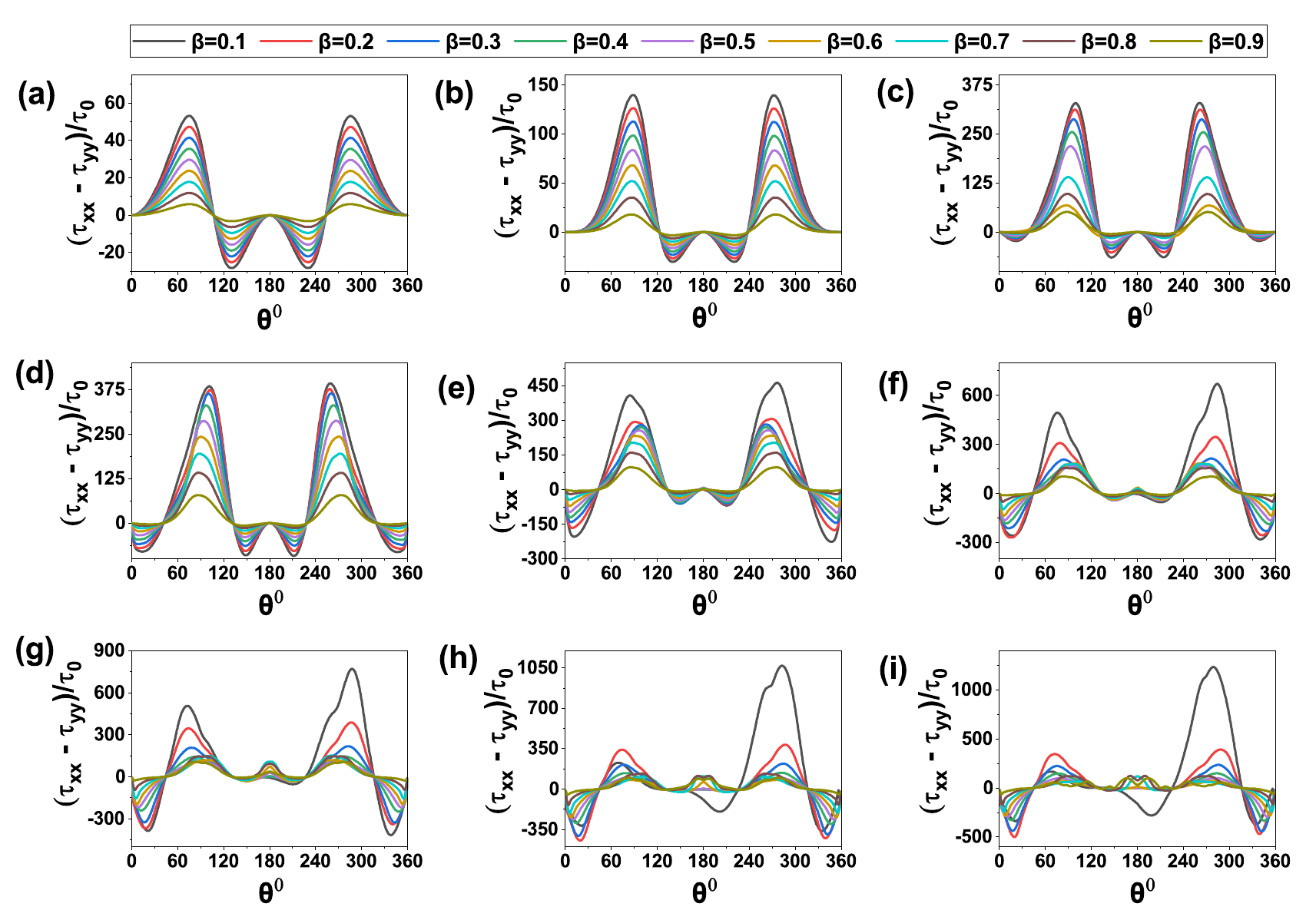}
\caption{Variation of first normal stress difference ($N_{1}/\tau_0$) over the surface ($0^\circ\le \theta\le 360^\circ$) of the cylinder cylinder for different $\beta$ at (a) $De=0.025$ (b) $De=0.1$ (c) $De=0.3$ (d) $De=0.5$ (e) $De=0.7$ (f) $De=0.9$ (g) $De=1.1$ (h) $De=1.3$ (i) $De=1.5$.} 
\label{stressdifftheta}
\end{figure}
\begin{figure}[!hbt]
	\centering
	\includegraphics[width=0.75\textwidth]{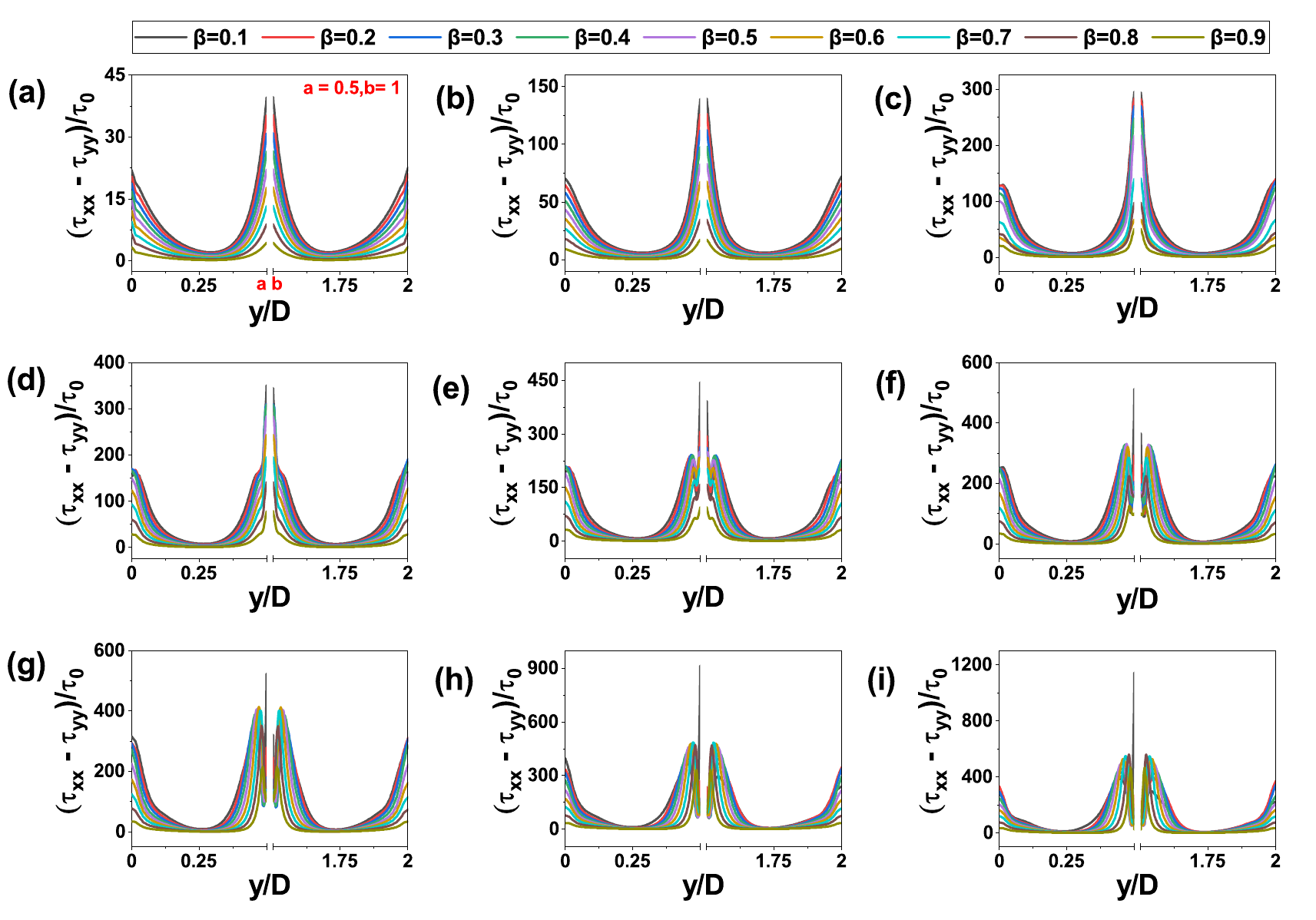}
	\caption{Variation of normal stress difference ($N_{1}/\tau_0$) along line A ($x_c, y$), (refer Fig. \ref{flow_domain}) for different $\beta$ at (a) $De=0.025$ (b) $De=0.1$ (c) $De=0.3$ (d) $De=0.5$ (e) $De=0.7$ (f) $De=0.9$ (g) $De=1.1$ (h) $De=1.3$ (i) $De=1.5$.} 
	\label{stressdiffline}
\end{figure}
\\
\noindent
Furthermore, the existing literature \citep{claus2013} suggests that the contribution of the flow-dependent normal stress is much higher than the contribution of the flow-dependent shear stress.  Fig. \ref{stressdifftheta} 
describes the variation of normal stress difference ($N_{1}/\tau_0$) over the cylinder surface.  A comparison of Figs. \ref{Wsstheta} and \ref{tauxxtheta} clearly state that the normal stress ($\tau_{xx}$) is much more dominated as compared to WSS ($\tau_{w}$). As the $De$ increases, a dip in the profiles at RSP ($\theta = 0^\circ$) clearly indicates the dominance of  $\tau_{yy}$ at that location. Also, at higher $De$, there is a flat profile with lower values of normal stress difference for lower $\beta$, which indicates that $\tau_{xx}$ and $\tau_{yy}$ components of stress are balanced at FSP. But at lower $\beta$, as the polymeric effect tends to dominate, there is great dominance of $\tau_{xx}$ component of stress.
\\
Further, Fig. \ref{stressdiffline} depicts the variation of the first normal stress difference ($N_{1}/\tau_0$) over the line A (refer Fig. \ref{flow_domain}) for the considered ranges of conditions ($\beta$, $De$). As expected, $N_{1}$ values are larger near the channel wall and adjacent to the cylinder surface on both the upper and lower sides of the cylinder, and the values reduce between the gap, due to the solid and stationary walls. The larger values are obtained adjacent to the cylinder surface at lower values of $\beta$ and $De$; however, increasing $De$ shows complex trends consistent with the above-discussed findings.
\begin{figure}[!b]
	\centering
	\includegraphics[width=\textwidth]{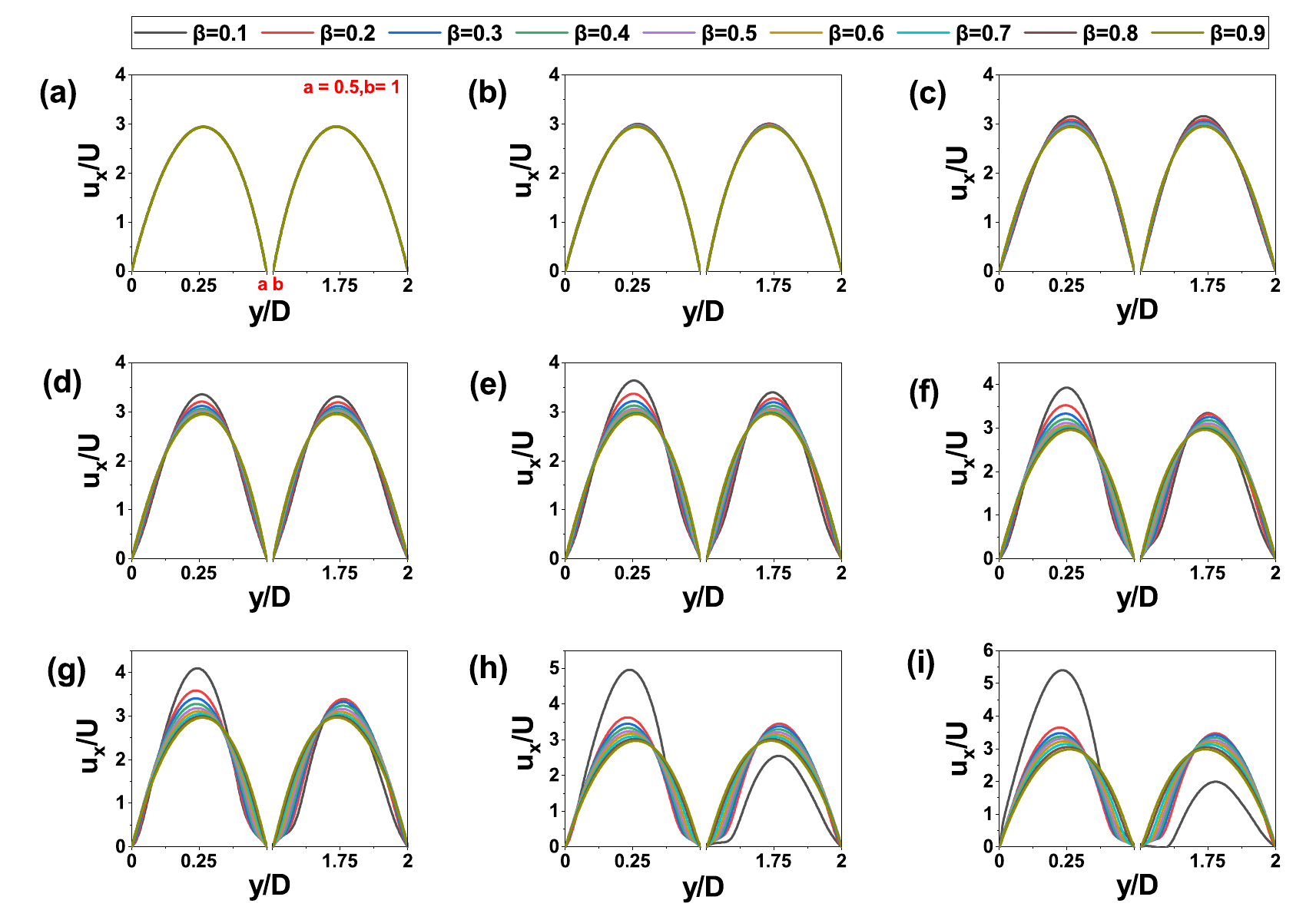}
	\caption{Velocity profile along line A ($x_c, y$), (refer Fig. \ref{flow_domain}) for different $\beta$ at (a) $De=0.025$ (b) $De=0.1$ (c) $De=0.3$ (d) $De=0.5$ (e) $De=0.7$ (f) $De=0.9$ (g) $De=1.1$ (h) $De=1.3$ (i) $De=1.5$.} 
	\label{velocityline}
\end{figure}
%
\subsection{Velocity}
%
\noindent 
Subsequently, Fig. \ref{velocityline} depicts the variation of the $x$-component of the velocity ($u_{\text{x}}/U_{\text{avg}}$) over the line A (refer Fig. \ref{flow_domain}) for the considered ranges of conditions ($\beta$, $De$). As both channel wall and cylinder are no-slip and impermeable, the velocity appears zero ($u_x = 0$) at these locations in Fig. \ref{velocityline}. Further, the similar profiles for lower $De\le 0.9$ in the upper and lower sides of the cylinder, refer Fig. \ref{velocityline}, deviate with increasing $De$. The velocity increases between the gap, in both upper and lower sides of the cylinder, due to decreasing flow area for fixed volumetric flow. The velocity profiles, therefore, appear parabolic, especially for lower $De$, with minor deviations at higher $De$. It is noteworthy evident through two peaks in each profile. The velocity remains unaffected by $\beta$ for low $De\le 0.3$. Nevertheless, at progressively higher Deborah numbers ($De$), lower $\beta$ values negatively affect become on the lower region of the cylinder compared to the upper region. Furthermore, the velocity in the lower gap is significantly reduced compared to the same in the upper gap. Also, the velocity gradient at the surface of the cylinder decreases for low $\beta$ values at higher $De$.   
%
\subsection{Drag Characteristics}
\begin{figure}[!b]
	\centering
	\includegraphics[width=\textwidth]{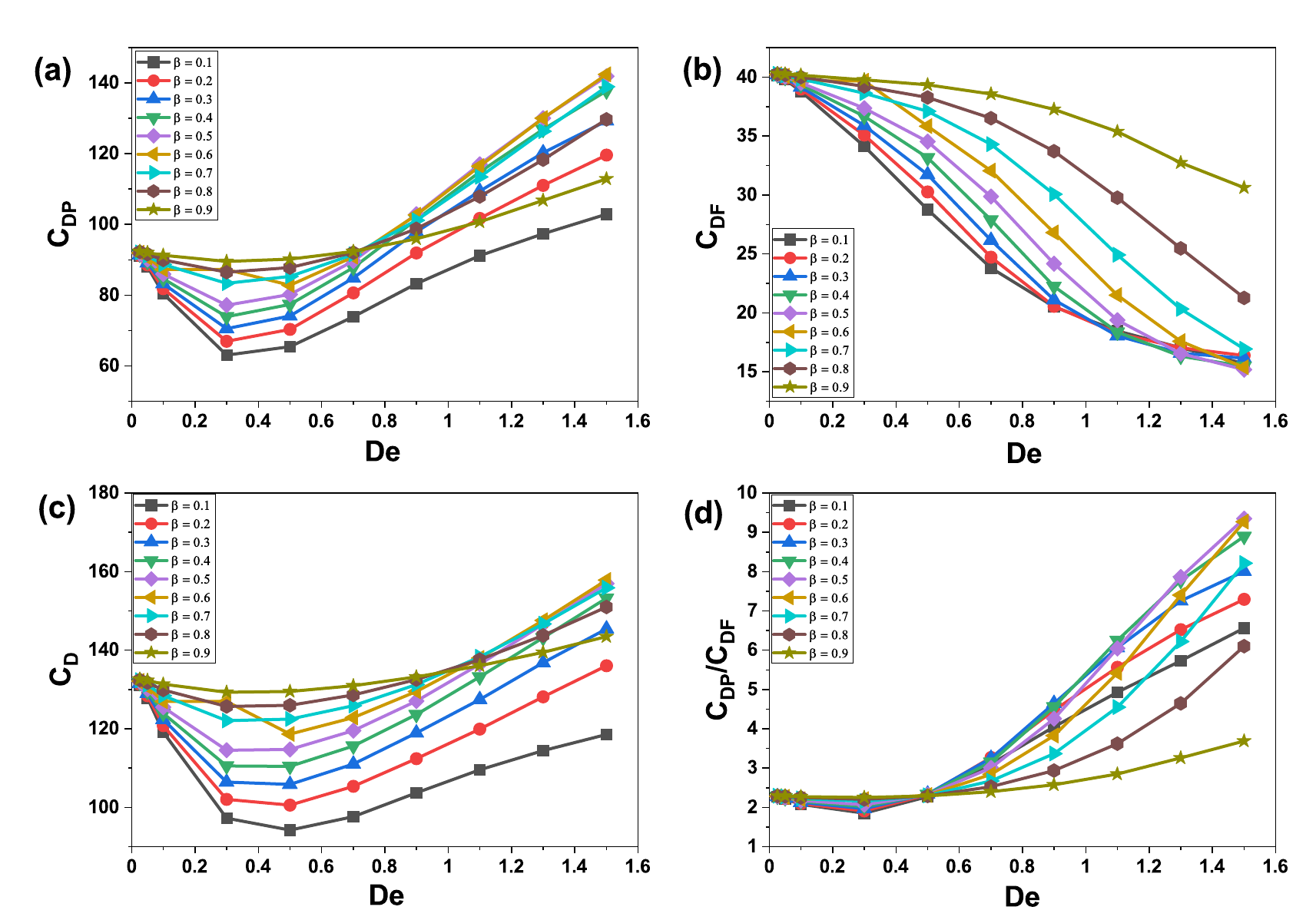}
	\caption{Dependence of the (a) pressure drag coefficient $C_{\text{DP}}$ (b) friction drag coefficient $C_{\text{DF}}$ (c) total drag coefficient $C_{\text{D}}$  (d) the drag ratio $C_{\text{DR}} = C_{\text{DP}}/C_{\text{DF}}$ on the Deborah number ($De$) and solvent viscosity ratio $(\beta)$.} 
	\label{Cd}
\end{figure}

\noindent 
The above sections have depicted the stronger role of flow governing parameters ($De$, $\beta$) of the local flow behavior, such as streamline and stress contour profiles, line profiles for the wall shear stress, pressure coefficient, normal stress, first normal stress difference, and velocity. Subsequently, the global engineering parameters, such as the drag coefficient, are analyzed in this section to understand the influence of these complex local flow characteristics manipulated by the governing parameters ($\beta$, $De$). In general, the total drag coefficient ($C_{\text{D}}$) is a contribution of the pressure drag coefficient  ($C_{\text{DP}}$), and the viscous drag coefficient  ($C_{\text{DF}}$), as defined in Eq. (\ref{eq:cd}). Figure \ref{Cd} illustrates the dependence of the drag coefficient and its component ($C_{\text{D}}$, $C_{\text{DP}}$, $C_{\text{DP}}$, $C_{\text{DP}}/C_{\text{DF}}$) on the Deborah number ($De$) and solvent viscosity ratio ($\beta$). 
%
%
\\\noindent
Fig. \ref{Cd}(a) shows the dependence of pressure drag coefficient ($C_{\text{DP}}$) on the governing parameters ($De$, $\beta$). For a fixed value of $\beta$, as the Deborah number increases gradually, initially, $C_{\text{DP}}$ decreases up to $De=0.3$, after that, it increases, and the increase is much more significant at higher $\beta$.  For lower values of $De$ ($\le 0.7$), the pressure drag coefficient ($C_{\text{DP}}$) decreases as  $\beta$ decreases. However, at higher values of $De$, the dependence of $C_{\text{DP}}$ on $\beta$ becomes rather complex; this may be attributed to as the relaxation time ($\lambda$) increases, viscoelastic effects enhance, and thus the polymeric nature complexly influence $C_{\text{DP}}$. Furthermore, at the higher value of Deborah number ($De$), there is a  strong dependence of $C_{\text{DP}}$ on the solvent viscosity ratio ($\beta$).   
%
\\\noindent
Fig. \ref{Cd}(b) shows the dependence of friction drag coefficient ($C_{\text{DF}}$) on the governing parameters ($De$, $\beta$). For a fixed value of Deborah number ($De$), the viscous drag coefficient ($C_{\text{DF}}$) decreases with a decrease in $\beta$.  For a fixed $\beta$, $C_{\text{DF}}$ decreases with an increase in $De$, as the relaxation time and polymeric effect increase, the fluid remains attached to the cylinder for a more extended period, thus lowering the $C_{\text{DF}}$. Further, at higher $De$, $C_{\text{DF}}$ values are almost equal for lower values of $\beta \le 0.7$. This phenomenon may be attributed to the increased polymeric effects with decreasing $\beta$. As discussed with $C_{\text{DP}}$, the fluid remains attached to the cylinder for an extended time. Hence, there is no significant change in $C_{\text{DF}}$ value. 
\\\noindent
Subsequently, Fig. \ref{Cd}(c) shows the dependence of total drag coefficient ($C_{\text{D}}$) on the  governing parameters ($De$, $\beta$). Qualitatively, the total drag coefficient ($C_{\text{D}}$) trends are similar to that of the pressure drag coefficient ($C_{\text{DP}}$). With increasing $De$, initially, there is a decrease in the value of $C_{\text{D}}$ {upto $De = 0.5$}{for $De \le 0.5$, and} then it starts to increase.  For lower values of $De\le 0.9$, as the value of $\beta$ increases, there is an increase in $C_{\text{D}}$ due to dominating $C_{\text{DP}}$.  For higher values of $De$, the dependence of $C_{\text{D}}$ upon $\beta$ is complex, but it still follows the trend at lower $\beta$.  At lower $De$, the value of $C_{\text{D}}$ is almost same, but for the moderate $De$ (say 0.7), there is a $34.04 \%$ increase in $C_D$ as the $\beta$ is increased from 0.1 to 0.9. However, at $De = 1.5$, there is a $20.99\%$ increase in $C_{\text{D}}$ as $\beta$ is varied from 0.1 to 0.9.
\\\noindent
Furthermore, to analyze the relative importance of the individual drag components, Fig. \ref{Cd}(d) shows the dependence of drag ratio ($C_{\text{DR}} =C_{\text{DP}}/C_{\text{DF}}$) on the governing parameters ($De$, $\beta$). As indicated from $C_{\text{DP}}$ and $C_{\text{DF}}$, the drag ratio appears to be greater than 1 (i.e., $C_{\text{DR}} > 1$) for the ranges of $De$ and $\beta$ suggesting the dominance of pressure forces in this flow. Further, the dominance of pressure drag is more pronounced at the higher Deborah number ($De$). Overall, the drag characteristics are complexly influenced by the flow governing parameters ($De$, $\beta$).
\section{Conclusions}
\noindent
In this study, the hydrodynamics of Oldroyd-B fluid flowing around a channel confined circular cylinder is investigated numerically. The numerical modeling and simulations are performed using the finite volume method open-source CFD solver, rheoTool based on OpenFOAM, for the following ranges of conditions: Deborah number ($0.025\leq De \leq1.5$), solvent viscosity ratio ($0.1\leq\beta\leq0.9$), blockage ratio ($B =0.5$) under the creeping flow (Reynolds number, $Re=0.01$) regime. The detailed kinematics in terms of the streamline, pressure and stress contour profiles, and engineering parameters as drag coefficient have been presented and discussed. Further detailed kinematics understandings are gained by analyzing the line plots for pressure coefficients, wall shear stress, normal stress, and normal stress difference over the surface of the cylinder and on the vertical line in the gap between the cylinder and the channel wall. For lower $De$, {we can see that the contours of} the flow field controur profiles  are perfectly symmetric for all values of $\beta$. However, at higher values of $De$, the flow contours show asymmetric behavior at lower $\beta$ where the polymeric effects are significant, indicating a difference in flow pattern at the top and bottom side of the cylinder.  The line plots in between the gap have shown the symmetric flow in both the top and bottom sides of the cylinder at lower $De$; the symmetry is lost with increasing $De$, particularly at lower $\beta$. Further, both $De$ and $\beta$ have a significant influence on the drag coefficient and its components ($C_{\text{DP}}$,$C_{\text{DF}}$, $C_{\text{D}}$).  At lower $De$, $C_{\text{D}}$ value is almost same irrespective of $\beta$ but {as we move to higher $De$ (=1.5)} there is a 20.99 $\%$ increment in $C_{\text{D}}$ value as $\beta$ is increased from 0.1 to 0.9 at higher $De$ (=1.5).  With an increase in $\beta$ at high $De$, the fluid transitions from elastic to viscous nature, leading to a subsequent rise in the drag coefficient. Over the range of parameters, the pressure drag force dominates over the frictional drag force in the flow. Overall, at lower $De$, the influence of $\beta$ is less prominent but at higher value of $De$, the influence of $\beta$ is significant.
\section*{CRediT Authors Contributions Statement}
%
\noindent 
In this work, the authors contributed as follows.
\begin{table*}[!h]
	\begin{tabular}{|c|p{0.28\linewidth}|p{0.54\linewidth}|}
		\hline
		Author No. & Author Name	& Contribution \\\hline
		1 & Pratyush Kumar {Mohanty}	&  Open-source Software, Investigation,  Data Curation, Validation, Visualization, Formal analysis, Writing - Original Draft\\\hline
		2 & Akhilesh Kumar {Sahu}	& Supervision, Conceptualization, Methodology, Resources, Formal analysis, Writing - Review \& Editing \\\hline
		3 & Ram Prakash {Bharti}	& Supervision, Conceptualization, Methodology, Resources, Open-source Software, Formal analysis, Writing - Review \& Editing \\\hline
	\end{tabular}
\end{table*}
%
\section*{Declaration of Competing Interest}
\noindent 
The authors declare that they have no known competing financial interests or personal relationships that could have appeared to influence the work reported in this paper.
%
\section*{Acknowledgements}
\noindent 
P.K. Mohanty would like to acknowledge the internship opportunity at the Complex Fluid Dynamics and Microfluidics (CFDM) Lab, Department of Chemical Engineering, Indian Institute of Technology Roorkee, India.
%
%
\nomenclature{$C_{\text{DF}}$}{Viscous component of drag coefficient, dimensionless}
\nomenclature{$C_{\text{DP}}$}{Pressure component of drag coefficient, dimensionless}
\nomenclature{$De$}{Deborah number, dimensionless}
\nomenclature{$B$}{Blockage ratio, dimensionless}
\nomenclature{$D$}{Diameter of the cylinder (m)}
\nomenclature{$C_p$}{Coefficient of pressure, dimensionless}
\nomenclature{$F_{\text{D}}$}{Drag force per unit length of the cylinder (N/m)}
\nomenclature{$C_{\text{D}}$}{Total drag coefficient, dimensionless}
\nomenclature{$H$}{Height of the computational domain, dimensionless}
\nomenclature{$I$}{Identity tensor, dimensionless}
\nomenclature{$L$}{Length of the computational domain, dimensionless}
\nomenclature{$L_{\text{d}}$}{Downstream length of the computational domain, dimensionless}
\nomenclature{$L_{\text{u}}$}{Upstream length of the computational domain, dimensionless}
\nomenclature{$p$}{Pressure (Pa)}
\nomenclature{$Re$}{Reynolds number, dimensionless}
\nomenclature{$u_x$}{Velocity in x-direction (m/s)}
\nomenclature{$u_y$}{Velocity in y-direction (m/s)}
\nomenclature{$U_{\text{avg}}$}{Average inlet velocity (m/s)}
\nomenclature{$x$}{Stream-wise coordinate (m)}
\nomenclature{$y$}{Transverse coordinate (m)}
\nomenclature[G]{$\beta$}{Solvent viscosity ratio, dimensionless}
\nomenclature[G]{$\eta_0$}{Total viscosity (Pa.s)}
\nomenclature[G]{$\eta_p$}{Polymeric viscosity (Pa.s)}
\nomenclature[G]{$\eta_s$}{Solvent viscosity (Pa.s)}
\nomenclature[G]{$\lambda$}{Relaxation time (s)}
\nomenclature[G]{$\rho$}{Density of the fluid (Kg/$m^3$)}
\nomenclature[G]{$\tau$}{Total extra-stress tensor (Pa)}
\nomenclature[G]{$\tau_p$}{Polymeric contribution in the extra-stress tensor (Pa)}
\nomenclature[G]{$\tau_s$}{Solvent contribution in the extra-stress tensor (Pa)}
\nomenclature[M]{CFD}{Computational Fluid Dynamics}
\nomenclature[M]{CUBISTA}{Convergent and Universally Bounded Interpolation Scheme for Treatment of Advection}
\nomenclature[M]{DAVSS}{Discrete Adaptive Elastic Viscous Split Stress}
\nomenclature[M]{DILU}{Diagonal based Incomplete LU}
\nomenclature[M]{PBiCG}{Preconditioned Biconjugate Gradient}
\nomenclature[M]{SIMPLE}{Semi Implicit Pressure Linked Equation}
\begin{spacing}{1.1}
\fontsize{11}{16pt}\selectfont
	\printnomenclature[5em]
\end{spacing}
\bibliography{references}



\end{document}